\documentclass[letterpaper,11pt]{article}

\usepackage{scrextend}
\usepackage{amsmath,amssymb,epsfig,bbm}
\usepackage{MnSymbol}
\usepackage{slashed}
\usepackage{amsthm}

\usepackage{float}

\usepackage{color}
\usepackage{mathrsfs}

\usepackage{dsfont}

\definecolor{co}{cmyk}{0,0.7,0.3,0}
\definecolor{darkgreen}{cmyk}{1,0,1,.2}
\definecolor{m}{rgb}{1,0.1,1}

\newcommand{\be}{\begin{equation}}
\newcommand{\ba}{\begin{eqnarray}}
\newcommand{\ea}{\end{eqnarray}}
\newcommand{\nn}{\nonumber}

\def\a{\alpha}
\def\b{\beta}

\def\d{\delta}

\def\m{\mu}
\def\n{\nu}
\def\oo{\omega}

\def\OO{\Omega}

\def\ca{{\cal A}}
\def\cb{{\cal B}}

\def\cf{{\cal F}}

\def\ch{{\cal H}}

\def\cl{{\cal L}}
\def\cm{{\cal M}}

\def\co{{\cal O}}

\makeatletter
\newcommand{\eqnum}{\refstepcounter{equation}\textup{\tagform@{\theequation}}}
\makeatother

\newcommand{\pa}{\partial}

\newtheorem{remark}{Remark}

\newtheorem*{definition*}{Definition}

\newcommand{\R}{\operatorname{\bold R}}

\newcommand{\Tr}{\operatorname{Tr}}

\fontfamily{yfrak}

\begin{document}

\vskip 25mm

\begin{center}

{\large\bfseries

The Metric Nature of Matter

}

\vskip 6ex

Johannes \textsc{Aastrup}$^{a}$\footnote{email: \texttt{aastrup@math.uni-hannover.de}} \&
Jesper M\o ller \textsc{Grimstrup}$^{b}$\footnote{email: \texttt{jesper.grimstrup@gmail.com}}\\ 
\vskip 3ex

$^{a}\,$\textit{Mathematisches Institut, Universit\"at Hannover, \\ Welfengarten 1, 
D-30167 Hannover, Germany.}
\\[3ex]
$^{b}\,$\textit{Independent researcher, Copenhagen, Denmark.}
\\[3ex]

{\footnotesize\it This work is financially supported by Ilyas Khan,  St.\\ Edmunds College, Cambridge, United Kingdom, by entrepreneur\\   Kasper Bloch Gevaldig, Copenhagen, Denmark, and by\\ \vspace{-0,1cm}Regnestuen Haukohl \& K\o ppen, Copenhagen, Denmark.}

\end{center}

\vskip 3ex

\begin{abstract}

\vspace{0.5cm}

We construct a metric structure on a configuration space of gauge connections and show that it naturally produces a candidate for a non-perturbative, 3+1 dimensional Yang-Mills-Dirac quantum field theory on a curved background. The metric structure is an infinite-dimensional Bott-Dirac operator and the fermionic sector of the emerging quantum field theory
is generated by the infinite-dimensional Clifford algebra required to construct this operator. The Bott-Dirac operator interacts with the $\mathbf{HD}(M)$ algebra, which is a non-commutative algebra generated by holonomy-diffeomorphisms on the underlying manifold, i.e. parallel-transforms along flows of vector fields. This algebra combined with the Bott-Dirac operator encode the canonical commutation and anti-commutation relations of the quantised bosonic and fermionic fields. The square of the Bott-Dirac operator produces both the Yang-Mills Hamilton operator and the Dirac Hamilton operator as well as a topological Yang-Mills term alongside higher-derivative terms and a metric invariant. 

\end{abstract}

\newpage
\tableofcontents

\section{Introduction}


One of the most important tasks in modern theoretical physics is to explain the mathematical structure of the standard model of particle physics.
This task can be divided into three fundamental questions:
\begin{enumerate}
\item
{\it What is (non-perturbative) quantum field theory?}
\item
{\it Why does the standard model have its particular algebraic structure?} 
\item
{\it What role does gravity play in the standard model?}
\end{enumerate}
The purpose of this paper is to present an algebraic framework that aims at answering these three questions. We show that a geometrical construction {\it over} a configuration space of gauge connections gives rise to a candidate for a non-perturbative, 3+1 dimensional quantum gauge theory coupled to a fermionic sector on a curved background. The construction, which employs the machinery of non-commutative geometry, produces in a semi-classical limit an algebraic structure similar to the almost-commutative algebra that Chamseddine and Connes identified as the structural backbone of the standard model of particle physics \cite{Connes:1996gi}-\cite{Chamseddine:2012sw}.

The basic idea is to base a fundamental theory on an elementary algebra combined with a metric principle and derive everything from that.  The algebra that we propose is the $\mathbf{HD}(M)$ algebra \cite{Aastrup:2012vq,AGnew}, which is generated by holonomy-diffeomorphisms -- that is, parallel transports along flows of vector fields -- which can be interpreted as operator valued functions on a configuration space $\ca$ of gauge connections. It is on this configuration space that we formulate a geometrical structure that interacts with the $\mathbf{HD}(M)$ algebra and hence forms a type of spectral triple {\it over} this configuration space.

The geometrical structure, that we introduce, is an infinite-dimensional Bott-Dirac operator. This operator is a variant of an operator, that we constructed in \cite{Aastrup:2019yui,Aastrup:2017atr} and is similar to an infinite-dimensional Bott-Dirac operator constructed by Higson and Kasparov in \cite{Higson}. 
The key feature of the Bott-Dirac operator is that its square produces the Hamilton operator of a Yang-Mills theory coupled to a fermionic sector as well as a topological Yang-Mills term together with higher-order derivative terms. The way the Dirac Hamiltonian emerges is through a commutator between the Bott-Dirac operator and a certain functional related to the ground state of the theory; hence the Dirac Hamiltonian can be understood as a quantum fluctuation of the bosonic theory.

Moreover, the canonical commutation relations and the canonical anti-commutation relations of a bosonic and fermionic quantum gauge theory are derived from the Bott-Dirac operator and its interaction with the $\mathbf{HD}(M)$ algebra. Also, since the representations of the the $\mathbf{HD}(M)$ algebra are in part labelled by the metric of the underlying manifold, the emergent quantum field theory will be formulated on a curved background. This background remains classical.

The key difference between the operator, that we construct in this paper, and the operators constructed in \cite{Aastrup:2019yui,Aastrup:2017atr}, is that the fermionic sector consist of spin-half fermions. A major shortcoming of the previous constructions is that the fermionic sector consisted of spin-one fermions; this has now been rectified.

This construction offers a completely novel interpretation of the role of fermions in quantum field theory. The anti-commutation relations (CAR) algebra emerges from the infinite-dimensional Clifford algebra, that is required to construct the Bott-Dirac operator, just as a $2^d$-dimensional Clifford algebra is used to construct a Dirac operator on a $d$-dimensional manifold, and hence the quantised fermions play an intrinsically geometrical role. The fermionic Fock space is, so to say, a storage room for geometrical data of the underlying configuration space of gauge connections. 

A key technical ingredient in this construction is a metric on the configuration space of gauge connections. This metric has the form of a Sobolev norm and works as an ultra-violet regularisation that dampens modes beyond a certain scale. An early version of this norm was presented and analysed in \cite{Aastrup:2017vrm}, where it was shown that it lead to a separable and strongly continuous representation of the $\mathbf{QHD}(M)$ algebra, which is the $\mathbf{HD}(M)$ algebra extended with translations. This representation did, however, not preserve the gauge symmetry. The solution to this problem is to introduce a modified, gauge covariant Sobolev norm \cite{Aastrup:2019yui}. Concretely, this means that all spatial derivatives are replaced with covariant derivatives. The result is a metric on the configuration space of connections that is covariant. 

This shift to a gauge covariant metric changes, however, the entire construction in a profound manner. Since the new metric is gauge covariant it depends on the gauge field, which means that it is an operator in the Hilbert space $L^2(\ca)$. This, in turn, means that it is dynamical.

The fact that the metric on the configuration space of connections is dynamical implies that the ultra-violet regularisation is also dynamical. Normally a regularisation is a technical artefact that must be removed in order to obtain physical quantities. To suggest that a regularisation should be physical would normally be impossible since there is no way to determine which regularisation one should pick: there are virtually countless ways to regularise. All this changes when the regularisation is dynamical. In that case there is no need to pick one particular regularisation (except for some initial data) since there is an evolution between them. 

Indeed, the ultra-violet regularisation that we encounter must be interpreted as an integral part of our framework. It is part of the metric data on the configuration space of gauge connections and thus not a mere computational tool. 

From a perturbative quantum field theoretical perspective the covariant regularisation will give rise to an infinite series of higher-derivative interactions where the specific form of the regularisation generates the corresponding coupling constants. They, too, are dynamical.

An interesting feature of the Bott-Dirac operator is that its square produces alongside the aforementioned Hamiltonians also a metric invariant that is related to the eta-invariant first formulated by Atiyah, Patodi and Singer \cite{Atiyah}-\cite{AtiyahIII}. Basically the invariant is the sum of all the eigenvalues of a spatial Dirac operator and thus it probes the asymmetry of its spectrum. We are able to establish the existence of this invariant as well as its independence of the choice of regularisation, but we do not know what physical meaning it has.

One of the most interesting attempts to answer the second question above is found in the work of Chamseddine and Connes \cite{Connes:1996gi}-\cite{Chamseddine:2012sw}, where they show that the standard model of particle physics fits remarkably well into the framework of noncommutative geometry \cite{ConnesBook,1414300}. The fact that the classical standard model coupled to general relativity can be formulated in terms of an almost-commutative spectral triple is highly non-trivial \cite{Krajewski:1996se}-\cite{Jureit:2009ye} and raises a number of questions, such as
\begin{itemize}
\item[-]
{\it where does the almost-commutative algebra originated from?}
\item[-]
{\it how does quantum field theory fit into the framework of non-commutative geometry?}
\end{itemize}
which are exactly the questions, that we attempt to answer in the present paper. What we propose is that the almost-commutative algebra in the standard model -- i.e. an algebra of the form $C^\infty(M)\otimes A_{\mbox{\tiny finite}}$, where $A_{\mbox{\tiny finite}}$ is a certain matrix algebra designed to produce the gauge structure of the standard model -- originates from an algebra generated by holonomy-diffeomorphisms and that the rich algebraic structure of the standard model emerges from a geometrical construction based on such an algebra.

We would like to emphasise that the approach, which we propose, differs decidedly from other rigorous approaches to quantum field theory such as algebraic quantum field theory \cite{Bogo,Brunetti} (see also \cite{Buchholz} for recent results) and axiomatic quantum field theory \cite{Osterwalder:1973dx}-\cite{Haag:1992hx} by not being based on a set of axioms such as locality, Lorentz invariance and causality, but rather on a simple algebraic principle. Whether or not the ensuing quantum field theory will be local or Lorentz invariant or satisfy causality is not yet known. The representation of the ${\mathbf{QHD}}(M)$ algebra is inherently non-local \cite{Aastrup:2019yui,Aastrup:2017vrm} and thus it may be that the Lorentz symmetry will be amended with a scale transformation. Indeed, since it is not known whether the Lorentz symmetry is an exact symmetry of Nature \cite{Jacobson:2004rj} and since it is in fact believed that locality is {\it not} realised in Nature\footnote{It is generally believed that this non-locality should emerge in the context of a theory of quantum gravity, but we see no reason why it cannot emerge from a fundamental theory of quantum fields; a theory in which gravity remains classical.} \cite{Doplicher:1994tu}, we do not believe that we should base our approach on these principles. They may not be solid.

Finally we would like to point out that the notion of a distance on a configuration space of gauge connections is not new but
was discussed already by Feynman \cite{Feynman:1981ss} and Singer \cite{Singer:1981xw} (see also \cite{Orland:1996hm} and references therein). What is new is our level of mathematical rigour and the employment of non-commutative geometry, the combination of which opens the door to a unifying picture that ties fermionic and bosonic quantum field theory together in a novel way.

\subsection{An outline of the main idea}

Let us begin with an outline of the central idea behind our construction. The following is merely a sketch of the analysis carried out in the subsequent sections.

We start with a configuration space $\ca$ of gauge connections and consider translations thereon. Two arbitrary connections $A$ and $A'$ always differ by a one-form $\omega$
$$
A' = A+\omega,
$$
which corresponds to a translation operator $U_\omega$
$$
U_\omega \xi(A) = \xi(A+\omega)
$$
on functions $\xi$ on $\ca$. 
Now, if we consider infinitesimal translations $\frac{\pa}{\pa A_i}$, where $\{A_i\}$ is an orthonormal basis of one-forms, then we can form a Bott-Dirac type operator on $\ca$ of the form 
$$
B= \sum_{i=1}^\infty \left( \bar{c}_i \frac{\pa}{\pa A_i} + {c}_i F_i \right),
$$
where $F_i$ is the curvature of $A_i$ and where $(c_i,\bar{c}_i)$ are elements in an infinite-dimensional Clifford algebra. The square of $B$ gives us then the Hamiltonian of a Yang-Mills theory coupled to a fermionic field\footnote{This model is in fact very similar to a model that Witten presented in section 2 in his paper on topological quantum field theory \cite{Witten:1988ze}. Witten discarded his model precisely because it involves spin-one fermions; precisely the problem that we solve in this paper. }
$$
B^2 = \sum_{i=1}^\infty \left( \left(\frac{\pa}{\pa A_i}\right)^2 +  \left( F_i\right)^2 \right) + "\mbox{fermionic terms}"
$$
in a form, which resembles an infinite-dimensional harmonic oscillator. Here the fermionic sector emerges from the infinite-dimensional Clifford algebra, which is required in order to construct $B$, i.e.
$$
\{ c_i , {c}_j \} = \d_{ij},\quad \{ \bar{c}_i , \bar{c}_j \} = -\d_{ij},\quad \{ c_i , \bar{c}_j \} = 0,
$$
gives rise to the canonical anti-commutation relations
$$
\{  {\psi}^\dagger({\bf m}_1) , {\psi}({\bf m}_2) \} = \d({\bf m}_1-{\bf m}_2).
$$
Furthermore, the state
$$
\Psi(A) = e^{i CS(A)}
$$
where
$$
CS(A) = \int_M \mbox{Tr} \left( A\wedge dA + \frac{2}{3} A\wedge A \wedge A \right)
$$
is the Chern-Simons functional, will lie in the kernel of $B$, i.e.
$$
B \Psi(A) =0,
$$
which thus gives us the ground state of the theory.

In \cite{Aastrup:2019yui} we rigorously developed this idea and found an infinite-dimensional Bott-Dirac operator, which we represented in a Hilbert space $L^2(\ca)\otimes \Lambda^* T\ca$ where the factor $\Lambda^* T\ca$ is the CAR algebra build from vectors in the tangent bundle over $\ca$. This CAR algebra then gave rise to the fermionic sector.

There is, however, a problem with this idea, which is that the fermions will a priori be one-forms, i.e. there will be a one-to-one correspondence between the bosonic and the fermionic sectors, which is at odds with special relativity and the spin-statistics theorem. 
The solution, which we present in this paper, is to redefine the Bott-Dirac operator as
$$
B= \sum_{i,j=1}^\infty \left( b_{ij}\bar{c}_i \frac{\pa}{\pa A_j} + b_{ij}{c}_i F_j\right),
$$
where $(c_i,\bar{c}_j)$ now belong to a CAR algebra build from spinors and where $b_{ij}$ are coefficients that satisfies a relation
$$
\sum_i b_{ij} b_{ik}^* \sim \d_{jk}.
$$
The square of the new Bott-Dirac operator then produces both the Yang-Mills and the Dirac Hamiltonians
$$
B^2= H_{\mbox{\tiny Yang-Mills}}\otimes \mathds{1} + H_{\mbox{\tiny Dirac}} \otimes \gamma + \ldots,
$$
where "$\ldots$" include a topological Yang-Mills terms, a metric invariant as well as higher-derivative terms.

The Bott-Dirac operator interacts with the ${\bf HD}(M)$ algebra generated by holonomy-diffeomorphisms and thus forms a type of noncommutative spectral triple
$$
(B, {\bf HD}(M), \ch )
$$
over $\ca$. Here $\ch$ is a Hilbert space that carries a representation of the ${\bf HD}(M)$ algebra.

\subsection{Notation and outline of the paper}

Throughout the paper $M$ denotes a 3-dimensional compact spin-$\mathbb{C}$ manifold with a trivial spin bundle\footnote{This assumption is probably too restrictive but for the time being we shall assume the existence of a trivial spin bundle in order to avoid the necessity of a local argument.}. We denote by $\{{\bf m},{\bf m}_1,{\bf m}_2 \ldots\}$ points in $M$ and by $\{m_\m\}$ a corresponding coordinate system where $\{\m,\n, \ldots\}$ are spatial indices. We denote by $(x_1,x_2,\ldots)$ coordinates in $\mathbb{R}^n$. Furthermore, we use $\{a,b,\ldots\}$ to index Lie-algebras and $\{i,j,\ldots\}$ to label orthonormal bases of $\OO^0(M,\mathfrak{g})$, $\OO^1(M,\mathfrak{g})$ and $L^2(M,\mathds{S}\otimes \mathfrak{g})$ where $\mathfrak{g}$ is a Lie algebra and $\mathds{S}$ is the spin bundle. Also, indices $({\a,\b, \ldots})$ label spin.
Throughout the paper we denote by $g$ a fixed metric on $M$ and by $e= e^a\sigma^a= e_\m^a\sigma^a dm^\m $ a compatible triad field, i.e. $g_{\m\n}= e^a_\m e^a_\n$, where repeated indices are summed.  
Also, we use the following summation convention: repeated $a,\mu$-type indices are summed, else all summations are explicitly written.\\

The paper is organised as follows: We begin in section 2 by introducing a metric on a configuration space of gauge connections, which we use to construct the infinite-dimensional Bott-Dirac operators in section 3. This construction is then doubled in section 4 where we obtain a setup that involves a ground state given by the Chern-Simons functional and a Bott-Dirac operator with a term linear in the field-strength tensor. In section 5 we demonstrate how the construction is related to bosonic and fermionic quantum field theory, in particular that the square of the Bott-Dirac operator gives the Hamilton operator of a Dirac-Yang-Mills theory. We then move on to consider local representation of the $\mathbf{HD}(M)$ and $\mathbf{QHD}(M)$ algebras in section 6. We end the paper in section 7 with a discussion.

\section{Metrics on configuration spaces of connections}
\label{ovn2}

Let $G$ be a compact Lie group, $V=M\times \mathbb{C}^j$ a vector-bundle over the manifold $M$ and $\ca$ the space of $G$-connections acting in $V$. Our aim is to construct a metric on $T\ca$, the tangent of $\ca$, and to do this we need to consider vector fields on $\ca$. Given an element $\xi\in \OO^1(M, \mathfrak{g})$, where $\mathfrak{g}$ is the Lie-algebra of $G$, we get a vector field on $\ca$ simply via 
\begin{equation}
\frac{\pa f}{\pa \xi} = \lim_{t\rightarrow 0} \frac{f(A + t\xi) - f(A)}{t}, \quad f\in C^\infty(\ca).
\label{mozart}
\end{equation}
In particular we can use this to identify $T_A\ca  $  with\footnote{In fact, a priory this only gives an embedding of $\Omega^1(M,\mathfrak{g})$ in $T_A\ca$. To make this point precise we need to describe a topology on $\ca$. In light of what we will do below it is natural to identify $\ca$ with $\Omega^1(M,\mathfrak{g})$ by choosing a fixed connection $A_0$. Then we can consider the scalar product (\ref{inner}) on $\ca$ for this fixed connection, and complete $\ca$ in this topology. This will render $\ca$ a Hilbert manifold, and as a Hilbert manifold the tangent spaces in a given point is isomorphic to the Hilbert space itself, i.e. we get the completion of $\Omega^1 (M,\mathfrak{g})$ with respect to the scalar product (\ref{inner}). Thus, with this topology $T_A\ca$ is a Sobolev completion of $\Omega^1 (M,\mathfrak{g})$.
Note that this construction is independent of the choice of $A_0$ since a different choice of $A_0$ will lead to an affine transformation combined with an equivalent norm.  } $\OO^1(M, \mathfrak{g})$.

Next we wish to construct a metric on $T_A\ca$. To do this we first consider the Hodge-Laplace operator. Given a metric $g$ on $M$ the operator reads 
$$
\Delta = d d^* + d^* d: \OO^k(M)\rightarrow \OO^k(M).
$$
We can extend the Hodge-Laplacian to $\OO^k(M,\mathfrak{g})$ by choosing an orthonormal basis of $\mathfrak{g}$. 
Note that: 
\begin{enumerate}
\item
the Hodge-Laplace operator is invariant under isometries,
\item
$\OO^1(M,\mathfrak{g})$ is a real vector space.
\end{enumerate}
We now  construct a covariant Hodge-Laplace operator 
\begin{eqnarray}
\Delta_{A} : \OO^k(M,\mathfrak{g})\rightarrow \OO^k(M,\mathfrak{g})
,\quad\Delta_{A} = (d+A) (d+A)^* + (d+A)^* (d+A),
\end{eqnarray}
which we restrict to $\Delta_{A}:\OO^1(M,\mathfrak{g})\rightarrow \OO^1(M,\mathfrak{g})$. $\Delta_A$ is simply the ordinary Hodge-Laplace operator where all derivatives are replaced with covariant derivatives. Finally we use this to define the metric on $T_A\ca$ 
\begin{equation}
\langle \xi \vert \eta \rangle_A := \int_M \left(  \left(  1 + \tau_1\Delta_{A}^p\right) \xi({\bf m})         , \left(  1 + \tau_1\Delta_{A}^p\right) \eta({\bf m})   \right)_{\mbox{\tiny 1-forms}}       dm,
\label{inner}
\end{equation}
where $\xi,\eta\in\OO^1(M,\mathfrak{g})$ and where $(\cdot,\cdot)_{\mbox{\tiny 1-forms}} $ denotes the point-wise scalar product on $\OO^1(M,\mathfrak{g})$ induced by $g$. Also $p$ and $\tau_1$ are real, positive parameters.
Note that the inner product (\ref{inner}) depends on $A$, i.e. for each $A$ we have a different inner product.
In \cite{Aastrup:2019yui} we showed that this metric on $\ca$ is gauge invariant.

Next we extend the above definitions to spinors $\psi_1,\psi_2$ in $L^2(M,\mathds{S}\otimes \mathfrak{g})$, where $\mathds{S}$ is the spin bundle, by writing  
\begin{equation}
\langle \psi_1 \vert \psi_2 \rangle_{A}^{\mbox{\tiny S}} := \int_M \left(  \left(  1 + \tau_1 (D^A) ^{2p}\right) \psi_1({\bf m})         , \left(  1 + \tau_1(D^A) ^{2p}\right) \psi_2({\bf m})   \right)_{\mbox{\tiny spinors}}        dm,
\label{innerspinor}
\end{equation}
where $(\cdot,\cdot)_{\mbox{\tiny spinors}} $ denotes the local scalar product in $S$ and where $D^A=i \sigma^a e^\m_a \nabla^{A,\oo}_\m$ is the spatial Dirac operator with $\nabla^{A,\oo}$ being the covariant derivative that includes the spin connection
$\oo$.

The definitions of inner products $\langle \cdot \vert \cdot \rangle_{A}$ and $\langle \cdot \vert \cdot \rangle_{A}^{\mbox{\tiny S}}$ have a natural generalization where we instead of the function $f(x)=( 1+ \tau_1 x^p)^{-1}$ take any bounded and nowhere vanishing   function $f:[0, \infty) \to \R$ with 
$$
\lim_{x\to \infty} f(x)=0.
$$ 
With this we define the inner products
\begin{equation}
\langle \xi \vert \eta \rangle_{A,f} := \int_M \left(  f^{-1}(\tau_1 \Delta_A) \xi({\bf m})         , f^{-1}(\tau_1 \Delta_A) \eta({\bf m})   \right)_{\mbox{\tiny 1-forms}}       dm,
\label{innerX}
\end{equation}
and 
\begin{equation}
\langle \psi_1 \vert \psi_2 \rangle_{A,f}^{\mbox{\tiny S}} := \int_M \left(  f^{-1}(\tau_1 (D^A)^2) \psi_1({\bf m})         ,  f^{-1}(\tau_1 (D^A)^2) \psi_2({\bf m})   \right)_{\mbox{\tiny spinors}}        dm.
\label{innerspinorX}
\end{equation}
In general we can say that a metric on $\ca$ consist of a metric on $M$ combined with a positive, real function $f$ on $\mathbb{R}_+$, where the latter encodes the scaling property of the metric on $\ca$ with respect to the manifold $M$.

The role of the inner products (\ref{inner})-(\ref{innerspinorX}) involving the Hodge-Laplace operator is to serve as an ultra-violet dampening by suppressing modes below the scale $\tau_1$. As we shall see later this covariant UV regularisation will, when seen from a perturbative point of view, give rise to an infinite number of couplings of ever-higher complexity where the function $f$ provides the coupling constant through its derivations at zero.
It is important to realise that as we in the next sections progress with the construction of our theory the metrics (\ref{inner}) and (\ref{innerX}) will be operators in the Hilbert space $L^{2}(\ca)$ and thus subject to time-evolution. 
If we instead of the covariant Hodge-Laplace operator define  (\ref{inner}) with the ordinary Hodge-Laplace operator, then the result will be the Sobolev norm, which we used in \cite{Aastrup:2017vrm}.
In that case it would be static (see section \ref{sec-time}).

\section{An infinite-dimensional Bott-Dirac operator}

In this section we construct the Bott-Dirac operator that plays the central role in our construction. We first construct an infinite-dimensional Clifford algebra and then combine this with a mapping between spinors and gauge fields, to construct the Bott-Dirac operator.

\subsection{Constructing the CAR algebra}

Once we have the inner product (\ref{innerspinor}) we can define the CAR bundle over the configuration space of spinors in $L^2(M,\mathds{S}\otimes \mathfrak{g})$ via the Fock space $\Lambda^* L^2(M,\mathds{S}\otimes \mathfrak{g})$. Denote by $\{\psi_i\}$ a basis in $L^2(M,\mathds{S}\otimes \mathfrak{g})$ that is orthonormal with respect to (\ref{innerspinor}).
Denote by $\mbox{ext}(\psi)$ the operator of external multiplication with $\psi\in L^2(M,\mathds{S}\otimes \mathfrak{g})$ on $\Lambda^*L^2(M,\mathds{S}\otimes \mathfrak{g})$, and denote by $\mbox{int}(\psi)$ its adjoint, i.e. the interior multiplication by $\psi$:
\begin{eqnarray}
\mbox{ext}(\psi) (\psi_1 \wedge  \ldots \wedge \psi_n) &=&  \psi\wedge \psi_1 \wedge  \ldots \wedge \psi_n,
\nn\\
\mbox{int} (\psi) (\psi_1 \wedge  \ldots \wedge \psi_n) &=& \sum_i (-1)^{i-1} \langle \psi, \psi_i \rangle_A^{\mbox{\tiny S}} \psi_1 \wedge \ldots \wedge \psi_{i-1} \wedge \psi_{i+1} \ldots \wedge \psi_n,
\nn
\end{eqnarray}
where $\psi, \psi_i\in L^2(M,S\otimes \mathfrak{g})$. We have the following relations:
\begin{eqnarray}
\{\mbox{ext}(\psi_1), \mbox{ext}(\psi_2)  \} &=& 0,
\nn\\
\{\mbox{int}(\psi_1), \mbox{int}(\psi_2)  \} &=& 0,
\nn\\
\{\mbox{ext}(\psi_1), \mbox{int}(\psi_2)  \} &=& \langle \psi_1, \psi_2 \rangle_A^{\mbox{\tiny S}}  
\end{eqnarray}
as well as
$$
\mbox{ext}(\psi)^* = \mbox{int}(\psi),\quad \mbox{int}(\psi)^* = \mbox{ext}(\psi),
$$
where $\{\cdot,\cdot\}$ is the anti-commutator. We define  the Clifford multiplication operators $\bar{c}_i$ and $c_i$ given by
\begin{eqnarray}
  c(\psi) &=& \mbox{ext}(\psi) + \mbox{int}(\psi),
\nn\\
 \bar{c}(\psi) &=& \mbox{ext}(\psi) - \mbox{int}(\psi) 
\end{eqnarray}
and 
\begin{eqnarray}
c_i =  c(\psi_i) 
,\quad
\bar{c}_i = \bar{c}(\psi_i) 
\end{eqnarray}
that satisfy the relations 
\begin{eqnarray}
 \{c_i, \bar{c}_j\} = 0, \quad
 \{c_i, c_j\} = \d_{ij}, \quad
 \{\bar{c}_i, \bar{c}_j\} =- \d_{ij},
 \label{mangec}
\end{eqnarray}
as well as
$$
c_i^*= c_i, \quad \bar{c}_i^* = - \bar{c}_i.
$$
Note here that 
$$
\mbox{ext}(i\psi)= i\mbox{ext}(\psi) ,\quad  \mbox{int}(i\psi) =  -  i\mbox{int}(\psi),
$$
which implies that
\begin{equation}
c(i\psi) = i \bar{c}(\psi), \quad \bar{c}(i\psi) = i c(\psi).
\label{signc}
\end{equation}
We shall also use the notation:
\begin{equation}
\left.
\begin{array}{c}
\mathfrak{a}^\dagger_i:= \mbox{ext}(\psi_i)
\\
\mathfrak{a}_i:=  \mbox{int}(\psi_i)
\end{array}
\right\}
\quad
\mbox{with}
\quad
\{\mathfrak{a}^\dagger_i,\mathfrak{a}_j\} = \d_{ij}.
\label{rolignu}
\end{equation}

Finally notice that since the inner product (\ref{innerspinor}) depends on $A$ so does the basis $\{\psi_i\}$ and hence also the Clifford algebra. This means that the commutators between elements of the Clifford algebra and vectors $\frac{\pa}{\pa \xi_i}$ do not vanish
\begin{equation}
\left[ \frac{\pa}{\pa \xi_i}, z \right]= \co(\tau_1) ,\quad  z  \in \{c_j, \bar{c}_j, \mathfrak{a}^\dagger_j, \mathfrak{a}_j  \ldots \}
\label{nonzero}
\end{equation}
but leave a contribution at the order of $\tau_1$. For instance
$$
\left[ \frac{\pa}{\pa \xi_i}, c_j \right] = \sum_k \left\langle \frac{\pa \psi_j }{\pa \xi_i} \vert \psi_k \right\rangle_{A}^{\mbox{\tiny S}}  c_k .
$$

\subsection{An infinite-dimensional Bott-Dirac operator}

We begin by noting that given a metric $g$ on $M$ there exist a canonical map from the co-tangent bundle into the Clifford algebra over $M$ 
$$
T^*M\rightarrow Cl(3), \quad v\rightarrow e^a(v)\sigma^a,
$$ 
where $e^a = e^a_\m \pa^\m$ is the triad field and where $v$ is a co-vector field. There exist two unitarily inequivalent representation of $Cl(3)$. If we choose one of them and if we pick an element $\chi \in C^\infty(M,\mathds{S})$ that lies in the kernel of $\nabla^\oo$, then we obtain the map
\begin{equation}
\varphi_\chi: T^*M \rightarrow \mathds{S},\quad \varphi_\chi(v) = e^a(v)\sigma^a \chi .
\end{equation}
In this way we get a map
$$
\varphi_\chi: L^2(M, T^*M)\rightarrow L^2(M,\mathds{S}),
$$
which straight forwardly extends to
$$
\varphi_\chi: \OO^1(M,\mathfrak{g})\rightarrow L^2(M,\mathds{S}\otimes\mathfrak{g}).
$$
We are going to use this map to construct the Bott-Dirac operator.
Denote again by $\{{\psi}_i\}$ a set of vectors in $L^2(M,\mathds{S}\otimes \mathfrak{g})$, which are orthonormal with respect to the inner product (\ref{innerspinor}) and by $\{c_i\}$ and $\{ \bar{c}_i \}$ the corresponding set of elements in the CAR algebra. Denote by $\{\xi_i\}$ a set of vectors in $\OO^1(M,\mathfrak{g})$, which are orthonormal with respect to (\ref{inner}).  
Given  $\chi \in C^\infty(M,\mathds{S})$ that lies in the kernel of $\nabla^\oo$ we define the infinite-dimensional Bott-Dirac operator
\begin{eqnarray}
B_\chi 
:=  \sum_{ij} b_{ij}(\chi)  \left(  \tau_2  \bar{c}_i      \frac{\pa}{\pa\xi_{j }} +    {c}_{i} \frac{\pa S}{\pa \xi_j}\right),
\label{F1}
\end{eqnarray}
where
\begin{eqnarray}
b_{ij} (\chi) :=  \langle {\varphi}_\chi(\xi_j), \psi_i \rangle_{A}^{\mbox{\tiny S}}
\label{b}
\end{eqnarray}
and where $S\in C^\infty(\ca)$ is a real functional that is bounded from below. The choice of $S(A)$, which is important for the construction of the Hilbert space in which $B_\chi$ acts, was discussed in detail in \cite{Aastrup:2019yui}. 

Note that $b_{ij}(\chi)$ is complex, which implies that $B_\chi$ is not self-adjoint.
Let us also  compute the square of $B_\chi$  
\begin{eqnarray}
\frac{1}{2}\left( B_\chi B_\chi^*+ B_\chi^* B_\chi \right)&=&  \sum_{ijk}  \frac{1}{2}\left(  b_{ij}  b^*_{ik} +  b^*_{ij}  b_{ik}  \right)  \left(  -\tau_2^2     \frac{\pa}{\pa\xi_{j }}  \frac{\pa}{\pa\xi_{k }}    
+  \frac{\pa S}{\pa \xi_j}\frac{\pa S}{\pa \xi_k}  \right)
\nn\\&&
+ \sum_{ijkl} \frac{\tau_2}{2} \left(  b_{ij} b_{kl}^*+ b_{ij}^* b_{kl} \right) \bar{c}_i   {c}_{k}   \frac{\pa^2 S}{\pa \xi_j\pa \xi_l }      
\nn\\&&
+ \sum_{ijkl} \frac{\tau_2}{2} \left(  b_{ij} b_{kl}^*+ b_{ij}^* b_{kl} \right) \bar{c}_i   \left[\frac{\pa}{\pa \xi_j }  , {c}_{k} \right]  \frac{\pa S}{\pa \xi_l }      
\nn\\&&
+ \sum_{ijkl} \frac{\left(\tau_2\right)^2}{2} \left(  b_{ij} b_{kl}^*+ b_{ij}^* b_{kl} \right) \bar{c}_i  \left[  \frac{\pa}{\pa \xi_j }  , \bar{c}_{k}\right]   \frac{\pa}{\pa \xi_l }  ,
\label{b2}
\end{eqnarray}
where the last two terms are due to (\ref{nonzero}) and thus of order $\tau_1$.

The Bott-Dirac operator in equation (\ref{F1}) depends on the spinor $\chi$. We assume that the spin bundle $\mathds{S}$ is trivial and chose two orthogonal spinors, that we use to construct two Bott-Dirac operators labelled by these spinors.  
Therefore let $\chi_\a\in L^2(M,\mathds{S})$, $\a\in\{1,2\}$, be a local orthonormal frame in the spin bundle $\mathds{S}$ over $M$, i.e. $(\chi_\a, \chi_\b)= \d_{\a\b}$, and denote by
\begin{equation}
B_\a := B_{\chi_\a}
\label{simpleminds}
\end{equation}
the corresponding set of Bott-Dirac operators. We still require $\chi_\a$ to lie in the kernel of $\nabla^\oo$. We also write
$$ b_{ij\a}:= b_{ij}(\chi_\a).$$
With this we compute the relations
\begin{eqnarray}
\sum_{i} b_{ij\a} b_{ik\b}^* &=&  \langle {\chi}_\a, \sigma^a \sigma^b e^a(\xi_j)  e^b(\xi_k)  \chi_\b \rangle_{A}^{\mbox{\tiny S}},
\label{K1}\\
\sum_{i\a} b_{ij\a} b_{ik\a}^*&=& \d_{jk}, 
\label{K3}\\
\sum_j b_{ij}(\chi)\xi_j &=& \left(\chi, \sigma^a   \psi_i  \right) e^a_\m  dm^\m ,
\label{K4}\\
\sum_j b^*_{ij}(\chi)\xi_j &=&- \left(   \psi_i, \sigma^a \chi   \right) e^a_\m  dm^\m ,
\label{K5}\\
\sum_i b_{ij}(\chi) \psi_i &=& \varphi_\chi(\xi_i),
\label{K6}
\end{eqnarray}
which will play a key role in the following. We define the operator
\begin{equation}
H:=\frac{1}{2}\sum_{\a=1}^2 \left( B_\a B_\a^*  +  B_\a^*B_\a  \right) 
\label{byraad}
\end{equation}
and compute
\begin{eqnarray}
H &=&  \sum_{i}    \left(  -\tau_2^2     \frac{\pa}{\pa\xi_{i }}  \frac{\pa}{\pa\xi_{i }}    
+  \frac{\pa S}{\pa \xi_i}\frac{\pa S}{\pa \xi_i}  \right)
\nn\\&&
+ \sum_{ijkl\a} \frac{\tau_2}{2} \left(  b_{ij\a} b_{kl\a}^*+ b_{ij\a}^* b_{kl\a} \right) \bar{c}_i   {c}_{k}    \frac{\pa^2 S}{\pa \xi_j\pa \xi_l } 
\nn\\&&
+ \sum_{ijkl\a} \frac{\tau_2}{2} \left(  b_{ij\a} b_{kl\a}^*+ b_{ij\a}^* b_{kl\a} \right) \bar{c}_i   \left[\frac{\pa}{\pa \xi_j }  , {c}_{k} \right]  \frac{\pa S}{\pa \xi_l }  
\nn\\&&
+ \sum_{ijkl\a} \frac{\left(\tau_2\right)^2}{2} \left(  b_{ij\a} b_{kl\a}^*+ b_{ij\a}^* b_{kl\a} \right) \bar{c}_i  \left[  \frac{\pa}{\pa \xi_j }  , \bar{c}_{k}\right]   \frac{\pa}{\pa \xi_l }     .
\label{b22}
\end{eqnarray}
Let us now assume that 
$$
S= \int_M \mbox{Tr}  P(F),
$$
where $P(F)$ is a polynomial in the curvature\footnote{This restriction on $P(F)$ can be relaxed but for now it suffices.} $F(A)$. With this assumption we find
\begin{eqnarray}
 \frac{\pa^2 S}{\pa \xi_j \pa \xi_l} = \int_M \left(\mbox{Tr} \left( \xi_{j}  \wedge Q   \xi_{l}   \right)   +   \left( \xi_{l}  \wedge Q  \xi_{j}   \right)\right),
\label{klipper}
\end{eqnarray}
where $Q=Q(F,\nabla^A)$ is an operator that is polynomial in $F(A)$ and the covariant derivative $\nabla^A$. With this we compute 
\begin{eqnarray}
-\sum_{ijkl\a} \frac{\tau_2}{2} \left(  b_{ij\a} b_{kl\a}^*+ b_{ij\a}^* b_{kl\a} \right)  \bar{c}_i   {c}_{k}   \frac{\pa^2 S}{\pa \xi_j\pa \xi_l }     \hspace{-5,5cm}&&\nn\\
&=&
\sum_{ik\a} \frac{\tau_2}{2}  ( \bar{c}_i   {c}_{k} +  \bar{c}_k  {c}_{i} )
 \int_M \left( 
  \mbox{Tr} \left(            (Q^{ab}\psi_k,  \sigma^b \chi_\a)     (\chi_\a, \sigma^a \psi_i)  \right) 
\right.\nn\\&&\left.
\hspace{3,8cm}+  \mbox{Tr} \left(         (\psi_k,  \sigma^a \chi_\a)   (\chi_\a, \sigma^b Q ^{ab} \psi_i)   \right)
 \right) ,
\label{startrek}
\end{eqnarray}
where $Q^{ab}= e^a_\m  dm^\m \wedge  Q   e^b_\n dm^\n$ and
where we used relation (\ref{K4}) together with the fact that $\chi_\a$ lies in the kernel of $\nabla^\oo$. We now use that
\begin{equation}
\sum_{\a=1}^2 \chi_\a )( \chi_\a = \mathds{1}_2
\label{coronaxxx}
\end{equation}
to finally obtain 
\begin{eqnarray}
H &=&  \sum_{i}    \left(  -\tau_2^2     \frac{\pa}{\pa\xi_{i }}  \frac{\pa}{\pa\xi_{i }}    
+  \frac{\pa S}{\pa \xi_i}\frac{\pa S}{\pa \xi_i}  \right)
\nn\\&&
- \sum_{ij} \frac{\tau_2}{2}      \int_M \left( \bar{c}_i   {c}_{j} + \bar{c}_j   {c}_{i}  \right)  
\left( \mbox{Tr} \left(    (Q^{ab}\psi_j,  \sigma^b \sigma^a \psi_i)  \right) 
\right. \nn\\ &&\left.\hspace{2cm}
+  \mbox{Tr} \left(     (\psi_j,  \sigma^a  \sigma^b Q ^{ab} \psi_i)   \right) \right)
+  \OO(\bar{c}, A),
\label{b222}
\end{eqnarray}
where
\begin{eqnarray}
\OO(\bar{c}, A) &=&
 \sum_{ijkl\a} \frac{\tau_2}{2} \left(  b_{ij\a} b_{kl\a}^*+ b_{ij\a}^* b_{kl\a} \right) \bar{c}_i   \left[\frac{\pa}{\pa \xi_j }  , {c}_{k} \right]  \frac{\pa S}{\pa \xi_l }  
\nn\\&&
+ \sum_{ijkl\a} \frac{\left(\tau_2\right)^2}{2} \left(  b_{ij\a} b_{kl\a}^*+ b_{ij}^* b_{kl} \right) \bar{c}_i  \left[  \frac{\pa}{\pa \xi_j }  , \bar{c}_{k}\right]   \frac{\pa}{\pa \xi_l }     ,
\label{notzero}
\end{eqnarray}
which proves that the operator $H$ is independent of the choice of orthonormal frame $\chi_\a$ in the limit $\tau_1=0$.\\

Note that the function 
\begin{equation}
 \Psi(A) = \exp\left(- \tau_2^{-1} S(A)  \right) \otimes 1 
\label{groundstate}
\end{equation}
in the Hilbert space $ L^2(\ca)\otimes \Lambda^* L^2(M,\mathds{S}\otimes\mathfrak{g})$, which we constructed in \cite{Aastrup:2019yui},
lies in the kernel of $B_\chi$, i.e.
\begin{equation}
B_\chi \Psi(A)=0,
\label{wakeup}
\end{equation}
where we used that $c_i 1 = \bar{c}_i 1 = \psi_i$. 
Relation (\ref{wakeup}) plays a key role when we construct the Hilbert space 
$$\ch=L^2(\ca)\otimes \Lambda^* L^2(M,\mathds{S}\otimes\mathfrak{g}),$$ which carries a representation of the $\mathbf{QHD}(M)$ algebra and in which the Bott-Dirac operator acts.
 The function $ \Psi(A)$ is the ground state. Since $ \Psi(A)$ is gauge invariant the construction of $L^2(\ca)$ involves a gauge fixing. Also, it is necessary to consider the subspace of $\Lambda^* L^2(M,\mathds{S}\otimes\mathfrak{g})$ that consist of the gauge covariant sections. We refer the reader to \cite{Aastrup:2019yui} for more information.

Note also that $B_\chi$ will be gauge covariant whenever $S(A)$ is gauge invariant.

Let us end this section with a few remarks:
\begin{remark}
Note that the construction of the Dirac operator (\ref{F1}) raises a question concerning
the expression $\frac{\pa}{\pa \xi_i}$, which requires a trivialisation of $T\ca$. We can either choose a global orthogonal frame, in which case (\ref{F1}) makes sense as it
stands albeit it depends on the global frame, or we can define 
the operator \cite{Aastrup:2019yui}
$$
B_\chi :=  \sum_{ij} b_{ij}(\chi)  \left(  \tau_2  \bar{c}_i     \nabla_{\xi_j}   +    {c}_{i} \frac{\pa S}{\pa \xi_j}\right),\qquad (\mbox{alternative def.})
$$
where $ \nabla_{\xi_i} $ denotes the Levi-Civita connection in $T\ca$. The Levi-Civita connection does indeed exist since $T\ca$ is a strong Riemannian manifold. The proof is the same as the proof in the finite case, see for example  \cite{michor}.

For the remainder of this paper we shall work with the first option, i.e. with the Bott-Dirac operator defined in (\ref{F1}) using a global trivialisation. In section \ref{secdiscussion} we shall then briefly discuss the possibility of working with a Levi-Civita connection in $T\ca$.
\end{remark}

\begin{remark}
The Bott-Dirac operator in (\ref{F1}) resembles the Bott-Dirac operator that Higson and Kasparov constructed in \cite{Higson}. If we in the definition of the metric (\ref{inner}) on $T_A\ca$ use the ordinary Hodge-Laplace operator $\Delta$ instead of the covariant Hodge-Laplace operator $\Delta_A$ and use the corresponding global orthonormal frame $\{\xi_i\}$ in $T\ca$ to construct the CAR algebra, then the Bott-Dirac operator would coincide with the one defined by Higson and Kasparov provided that we choose $S(A) =\tau_2 \int_M \vert A\vert ^2$ (for details see \cite{Aastrup:2017atr}). This Bott-Dirac operator does not depend on $\chi$ and its square coincides with the Hamilton operator of an infinite-dimensional harmonic oscillator. In \cite{Aastrup:2017atr} we showed that the square of such a Bott-Dirac operator is identical to the Hamilton operator of a free sector of a gauge field coupled to a fermionic vector-field.
\end{remark}

\begin{remark}

The construction of the Bott-Dirac operator in (\ref{F1}) relies on a pairing of spinors and vectors encoded in the matrix $b_{ij}(\chi)$. This pairing can, however, also be built directly into the Clifford algebra, which leads to an alternative Bott-Dirac operator 
$$
B'_\chi = \sum_i \left( \tau_2 \bar{c} (\varphi_\chi(\xi_i))\frac{\pa}{\pa \xi_i}   + {c}(\varphi_\chi(\xi_i)) \frac{\pa S}{\pa \xi_i} \right)
$$
 that is different from (\ref{F1}).
To see why $B'_\chi$ differ from $B_\chi$ we note that
$
\sum_i \psi_i b_{ij} = \varphi_\chi(\xi_j)
$, see (\ref{K6}),
does not imply that
\begin{equation}
c(\varphi(\xi_j)) = \sum_i c_i b_{ij}, \quad \bar{c}(\varphi(\xi_j)) = \sum_i \bar{c}_i b_{ij}, \quad (\mbox{wrong})
\label{wrong}
\end{equation}
which would imply that $B_\chi=B'_\chi$. The reason why (\ref{wrong}) does not always hold is that the $b_{ij}(\chi)$'s may be complex, in which case the $c_i$ and $\bar{c}_i$ sectors will mix according to (\ref{signc}). We shall not further discuss this option but instead proceed with the operator defined in (\ref{F1}).

\end{remark}

\subsection{Creation and annihilation operators}

It is possible to formulate the Bott-Dirac operator in terms of creation and annihilation operators. This is similar to what was done in section 2 in \cite{Aastrup:2017atr}. This formulation will not play a role in the following sections but we add it nevertheless. 

We define the operators 
\begin{eqnarray}
\mathfrak{q}_i := \left( \frac{\pa S}{\pa \xi_i}+ \tau_2  \frac{\pa}{\pa\xi_i}\right)  , &&\quad
\mathfrak{q}_i^\dagger :=  \left( \frac{\pa S}{\pa \xi_i} -\tau_2 \frac{\pa}{\pa\xi_i} \right) ,
\label{morten}
\end{eqnarray}
with the reverse
$$
 \frac{\pa S}{\pa \xi_i} =  \frac{1}{2}  (\mathfrak{q}_i +\mathfrak{q}_i^\dagger ) ,   \quad    \tau_2 \frac{\pa}{\pa\xi_i} =\frac{1}{2}  (\mathfrak{q}_i -\mathfrak{q}_i^\dagger ).
$$
The $\mathfrak{q}$ and $\mathfrak{q}^\dagger$ operators satisfy the relations  
\begin{eqnarray}
\left[\mathfrak{q}_i , \mathfrak{q}_j^\dagger \right] =  {2\tau_2} \frac{\pa^2 S}{\pa\xi_i \pa \xi_j} ,
&&
\left[\mathfrak{q}_i  , \mathfrak{q}_j\right] =  \left[\mathfrak{q}_i^\dagger  , \mathfrak{q}_j^\dagger\right] =   0,
\label{covi}
\end{eqnarray}
which would be the commutation relations for the infinite-dimensional harmonic oscillator if $ \frac{\pa^2 S}{\pa\xi_k \pa \xi_l}\propto \d_{kl}$ (in \cite{Aastrup:2017atr} we show that the CCR algebra emerges from such a setup). 
Using definition (\ref{rolignu}) this gives us
$$
B_\a =  \sum_{ij} b_{ij\a} \left(  \mathfrak{a}_i^\dagger  \mathfrak{q}_j +   \mathfrak{a}_i \mathfrak{q}_j^\dagger   \right)
$$
as well as 
\begin{eqnarray}
H= 
 \sum_{i} \mathfrak{q}_i^\dagger   \mathfrak{q}_i 
 +{\tau_2}\sum_{ijkl\a}  \left( \mathfrak{a}_i^\dagger  \mathfrak{a}_j  \left( b_{ik\a}b^*_{jl\a}  +b^*_{ik\a}b_{jl\a}   \right)    \right) \frac{\pa^2 S}{\pa\xi_k \pa \xi_l} +  \OO(\bar{c}, A)  .
\label{tja}
\end{eqnarray}
We see that if $\frac{\pa^2 S}{\pa\xi_k \pa \xi_l}  \propto \d_{kl}$ and if we take the flat limit $M\rightarrow \mathbb{R}^3$ then this would almost coincide with the Hamilton operator of a free gauge field coupled to a $\mathfrak{g}$-valued fermionic field. The difference is the absence of the momenta $\vert \vec{p} \vert$. The reason for this is that in the flat limit the definition of the creation and annihilation operators in (\ref{morten}) will not coincide with the creation and annihilation operators used in a canonical quantization procedure in quantum field theory. This can be remedied by changing the definition of the creation and annihilation operators to
\begin{eqnarray}
\tilde{\mathfrak{q}}_i =  \frac{1}{(\lambda_i)^{{1}/{4}}} \left( \frac{\pa S}{\pa \xi_i}+ {\tau_2}   \frac{\pa}{\pa\xi_i}\right)  , &&\quad
\tilde{\mathfrak{q}}_i^\dagger =  \frac{1}{(\lambda_i)^{{1}/{4}}}  \left( \frac{\pa S}{\pa \xi_i} - {\tau_2}\frac{\pa}{\pa\xi_i} \right) ,
\label{morten2}
\end{eqnarray}
where $\lambda_i$ are the eigenvalues of the Hodge-Laplace operator, i.e. $\Delta_A \xi_i=\lambda_i \xi$. The eigenvalues $\lambda_i$  will, however, in the general case depend on $A$, which will render the modified commutation relations (\ref{covi}) considerable more complicated. Since this shall not play a role in the following analysis we leave this question here.

\section{The Chern-Simons functional}
\label{secCS}

The square of the Bott-Dirac operator in equation (\ref{b222}) involves the term $\sum_i \left( \frac{\pa S}{\pa \xi_i}\right)^2$. As we discussed in \cite{Aastrup:2019yui} $\frac{\pa S}{\pa \xi_i}$ will not involve terms linear in the curvature $F$ when $S(A)$ is build of terms, which are quadratic in $F$ (see section 6 in \cite{Aastrup:2019yui}). It is, however. important that $\frac{\pa S}{\pa \xi_i}$ involves such a linear term since this will give the operator $H$ in (\ref{b222}) a term quadratic in the curvature and a fermionic term linear in the covariant derivative. This, in turn, will imply that $H$ includes the Yang-Mills and Dirac Hamilton operators. In this section, which is based on the framework developed in \cite{Aastrup:2019yui}, we shall therefore discuss how this may be achieved by introducing a complex phase.\\

We begin with the observation that if we add a Chern-Simons term 
\begin{equation}
CS(A) =  \int_M \mbox{Tr} \left( {A}\wedge d{A} + \frac{2}{3} {A}\wedge {A} \wedge {A}\  \right)
\label{CS}
\end{equation}
 to $S(A)$ as a complex phase
\begin{equation}
S'(A) =S(A) + \frac{{i}}{2} CS(A)
\label{first}
\end{equation}
and use that
\begin{equation}
\frac{\pa CS}{\pa \xi_i} =  2 \int_M \mbox{Tr}\left(\xi_i\wedge F(A)  \right)
\label{godhavn}
\end{equation}
then it gives us a term in $\frac{\pa S}{\pa \xi_i}$ that is linear in $F(A)$. The addition of a complex phase to the function 
\begin{equation}
\Psi'(A) =  \exp\left( - \tau_2^{-1} S'(A)  \right)
\label{groundstate2}
\end{equation}
in (\ref{groundstate}) raises, however, the problem that the Bott-Dirac operator $B_\a$ and its conjugate $B_\a^*$ will not have the same kernel. With definition (\ref{first}) the function $\Psi'(A)$ will lie in the kernel of $B_\a$ but not in the kernel of $B_\a^*$.

To solve this problem we first double the construction 
\begin{eqnarray}
\mathbf{HD}(M) \longrightarrow \mathbf{HD}^c(M) =  \left(\mathbf{HD}(M)\right)^{ \oplus 2}
\label{HDC}
\end{eqnarray}
where $\mathbf{HD}(M) $ is the algebra generated by holonomy-diffeomorphisms defined in \cite{Aastrup:2012vq,AGnew,Aastrup:2014ppa}, with
\begin{eqnarray}
\ch \longrightarrow \ch^c &=& \left( L^2(\ca)   \otimes  \Lambda^* L^2(M,\mathds{S}\otimes\mathfrak{g}) \right)^{\oplus 2}
\nn\\
&=& \ch_\circ \oplus \ch_\bullet,
\label{HC}
\end{eqnarray}
where we choose a representation of the CAR algebra so that 
 the ground state $\vert 0 \rangle$ in the Hilbert spaces $\ch_\circ$ and $\ch_\bullet$ are annihilated by the annihilation or creation operators respectively. If we denote the ground state in $\ch_\circ$ and $\ch_\bullet$ by $\vert 0 \rangle_\circ$ and $\vert 0 \rangle_\bullet$ respectively then this means that
 \begin{equation}
 \mathfrak{a}_i\vert 0 \rangle_\circ=0
 \quad
 \mbox{and} 
 \quad
 \mathfrak{a}_i^\dagger\vert 0 \rangle_\bullet=0\quad \forall i.
 \label{Miami}
 \end{equation}
Next we introduce the operators 
\begin{align}
B_{ij} &:= \bar{c}_i \frac{\pa}{\pa \xi_j}  + c_i  \frac{\pa S }{\pa\xi_{j }},
\nn\\
 C_{ij}  &:=       \mathfrak{a}_i^\dagger  \frac{1}{2}   \frac{\pa CS }{\pa\xi_{j }}  ,
\end{align}
which we use to define the self-adjoint operator
\begin{eqnarray}
\cb_{ij} :=   \left(
\begin{array}{cc}
 B_{ij}  & C_{ij} \\ 
C^*_{ij}    &  B_{ij}
\end{array}
\right).
\label{BDx2}
\end{eqnarray}
A short computation shows that the state 
\begin{equation}
\ch^c \ni \Phi (A) := \left(
\begin{array}{r}
\cos \left(({2\tau_2})^{-1}   CS(A)\right) \exp(-\tau_2^{-1}S(A))  
\vspace{0,1cm}
\\
 \sin\left(({2\tau_2})^{-1}  CS(A)\right)  \exp(-\tau_2^{-1}S(A)) 
\end{array}
\right) \otimes 1
\label{platter}
\end{equation}
lies in the kernel of $\cb_{ij}$, i.e.
\begin{equation}
\cb_{ij} \Phi(A) = 0.
\label{T1}
\end{equation}
Next we define the operator
\begin{eqnarray}
\cb_{\a}:= \sum_{ij} b_{ij\a} \cb_{ij} =   \left(
\begin{array}{cc}
 B_\a  & C_\a \\ 
\tilde{C}_\a    &  B_\a
\end{array}
\right),
\label{BDx22}
\end{eqnarray}
where
$$
C_\a := \sum_{ij} b_{ij\a} C_{ij} ,  \qquad     \tilde{C}_\a  := \sum_{ij} b_{ij\a} C_{ij}^*.
$$
The operator $\cb_\a$ is not self-adjoint since the coefficients $b_{ij\a}$ are not necessarily real. The functional $\Phi(A)$ lies nevertheless in the kernel of both $\cb_\a$ and its adjoint due to (\ref{T1}), i.e. 
$$
\cb_\a \Phi(A) = \cb^*_\a \Phi(A)  =0.
$$

\begin{remark}
Let us point out the similarity between $\Phi(A)$ in (\ref{platter}) and the Kodama state in canonical quantum gravity \cite{Kodama:1988yf} (see also \cite{Smolin:2002sz} and references therein). If we choose $\ca$ as a configuration space of Ashtekar connections \cite{Ashtekar:1986yd,Ashtekar:1987gu}, then $\Phi(A)$ would strongly resemble the Kodama state, albeit the present setup is different from the one presented in \cite{Smolin:2002sz}.
\end{remark}

\begin{remark}
We would like to stress the importance of involving the two different representations of the CAR algebra shown in (\ref{Miami}). First of all and as already mentioned this is critical in order to obtain relation (\ref{T1}). But equally important is the fact that if we had instead opted for a construction that involved only one of the representations of the CAR algebra then the Hamilton operator ${\bf H}$ -- or some variant thereof -- would involve not only commutators between the derivative $\frac{\pa}{\pa\xi_i}$ and either $\frac{\pa S}{\pa\xi_i}$ or $\frac{\pa CS}{\pa\xi_i}$ but also anti-commutators, which in turn would lead to highly non-local interactions in the ensuing quantum field theory\footnote{By highly non-local we mean terms, which are not even local in a perturbative sense.}. 
\end{remark}

Next we wish to compute the square of $\cb_\a$. We write 
\begin{eqnarray}
{\bf H}:= \frac{1}{2}\sum_\a \left( \cb_\a \cb_\a^*  + \cb_\a^* \cb_\a\right)  =
\left(
\begin{array}{cc}
H_{1}  & H^*_{2} \\
{H}_2 &  {H}_{1} 
\end{array}
\right) +  \OO'(\bar{c}, A)
\label{buddha}
\end{eqnarray}
with
\begin{align}
H_1 &= H + \frac{1}{4}  \sum_i \left( \frac{\pa CS}{\pa\xi_i} \right)^2
\nn\\
H_2^* &=  \frac{1}{2} \sum_\a  \left(  \left\{  C_\a   ,  B^*_\a    \right\}   +     \left\{  \tilde{C}_\a^*    ,   B_\a    \right\}     \right),
\end{align}
where $H$ is defined in (\ref{byraad}). Also, $ \OO'(\bar{c}, A)$ in (\ref{buddha}) is a term of order $\tau_1$ similar to (\ref{notzero}) that stems from a non-vanishing commutator between the vectors $\frac{\pa}{\pa\xi_i}$ and elements in the Clifford algebra.

 In total $H_1$ reads 
\begin{align}
H_1 &=  \sum_{i}    \left(  -\tau_2^2    \left( \frac{\pa}{\pa\xi_{i }}\right)^2     
+ \frac{1}{4}  \left( \frac{\pa CS}{\pa\xi_i} \right)^2  + \left( \frac{\pa S}{\pa \xi_i}\right)^2  \right)
\nn\\&
- \sum_{ij} \frac{\tau_2}{2}      \int_M \left( \bar{c}_i   {c}_{j} + \bar{c}_j   {c}_{i}  \right)  
\left( \mbox{Tr} \left(    (Q^{ab}\psi_j,  \sigma^b \sigma^a \psi_i)  \right) 
\right. \nn\\ &\left.\hspace{4cm}
+  \mbox{Tr} \left(     (\psi_j,  \sigma^a  \sigma^b Q ^{ab} \psi_i)   \right) \right).
\label{NewYork}
\end{align}
It is natural to split $H_1$ up in a bosonic and fermionic part
\begin{equation}
H_1 = H_1 \big\vert_{\mbox{\tiny bosonic}} + H_1\big\vert_{\mbox{\tiny fermionic}},
\label{Pl1}
\end{equation}
where the latter includes the terms, that involve the Clifford algebra elements.

Turning to $H_2$ we compute 
\begin{align}
H_2^*  &= \frac{1}{4} \sum_{ijkl\a}  ( b^*_{ij\a} b_{kl\a}  +   b_{ij\a} b^*_{kl\a}  )\left(    \mathfrak{a}^\dagger_i   \bar{c}_k  \frac{\pa CS}{\pa \xi_j}   \frac{\pa }{\pa \xi_l} +         \bar{c}_k \mathfrak{a}^\dagger_i   \frac{\pa }{\pa \xi_l}      \frac{\pa CS}{\pa \xi_j}                      \right)
\nn\\
&= 
 \frac{1}{4} \sum_{ijkl\a}  ( b^*_{ij\a} b_{kl\a}  +   b_{ij\a} b^*_{kl\a}  ) \left( 
- \bar{c}_k \mathfrak{a}^\dagger_i   \left[   \frac{\pa }{\pa \xi_l}    ,    \frac{\pa CS}{\pa \xi_j}      \right]  
- \d_{ki} \frac{\pa CS}{\pa \xi_j}   \frac{\pa }{\pa \xi_l} 
 \right)
 \nn\\
 &=
  \frac{1}{4} \sum_{ijkl\a}  ( b^*_{ij\a} b_{kl\a}  +   b_{ij\a} b^*_{kl\a}  ) \left(   
    \mathfrak{a}^\dagger_k \mathfrak{a}^\dagger_i      
  -  \mathfrak{a}_k \mathfrak{a}^\dagger_i     
\right)   \frac{\pa^2 CS}{\pa \xi_l \pa \xi_j}  
-\frac{1}{2} \sum_j \frac{\pa CS}{\pa \xi_j}   \frac{\pa }{\pa \xi_j} ,
\label{stiglitz}
\end{align}
where we used relation (\ref{K3}). 
Again it is natural to split $H_2$ up in a bosonic and fermionic part
\begin{equation}
H_2 = H_2\big\vert_{\mbox{\tiny bosonic}}   +   H_2\big\vert_{\mbox{\tiny fermionic}},
\label{Pl2}
\end{equation}
where the bosonic part consist of the term $-\frac{1}{2}  \sum_j \frac{\pa CS}{\pa \xi_j}   \frac{\pa }{\pa \xi_j} $ and where the fermionic part is the rest.

Let us move on and compute the fermionic part of $H_2$. A somewhat lengthy but straight forward computation gives us 
\begin{eqnarray}
-\sum_{jl\a} \frac{\pa^2 CS(A)}{\pa \xi_j \pa \xi_l}  \left(b^*_{ij\a} b_{kl\a} +  b_{ij\a} b^*_{kl\a} \right)
\hspace{-5cm}&&\nn\\
&{=}&
 \int_M \big( 
  \mbox{Tr} \left(            2i  ( \nabla^{A,\oo}_\a \psi_k, \sigma^a e_a^\a  \psi_i)    \right)   
-  \mbox{Tr} \left(      2i   (\psi_k, \sigma^a e_a^\a \nabla^{A,\oo}_\a \psi_i)   \right)
    \big)d^{(3)}\mbox{vol}
 \nn\\
&&\hspace{1,2cm} +  (k \leftrightarrow i),
\label{corona}
\end{eqnarray}
where we used 
$$
 \frac{\pa^2 CS}{\pa \xi_{i } \pa \xi_{j }} 
 =    \int_M \mbox{Tr} \left(  \xi_{i  }  \wedge \nabla^A  \xi_{j} \right) + \int_M \mbox{Tr} \left(  \xi_{j  }  \wedge \nabla^A  \xi_{i} \right)   ,   
$$
together with relation (\ref{K4}). Also, in (\ref{corona}) $\nabla^{A,\oo}= d+\oo + A$ where $\oo$ is the Levi-Civita spin connection. 
We note that the term in (\ref{stiglitz}) that is proportional to $ \mathfrak{a}^\dagger_k \mathfrak{a}^\dagger_i $ will, using (\ref{corona}), vanish since it is symmetric in the indices $i$ and $k$.

Finally we obtain  
\begin{equation}
H_2\big\vert_{\mbox{\tiny fermionic}} =   -\int_M  \mbox{Tr} \left(  \hat{\psi}^\dagger D^A \hat{\psi}\right)  + \int_M  \mbox{Tr} \left(  \hat{\psi}   D^A \hat{\psi}^\dagger\right) +  \mbox{Tr}_\psi\left( D^A  \right) ,
\label{opera}
\end{equation}
where $D^A= i \sigma^a e_a^\m \nabla_\m^{A,\oo}$ is a spatial Dirac operator and where
\begin{equation}
\hat{\psi}^\dagger  ({\bf m})  = \sum_i \mathfrak{a}_i^\dagger {\psi}^*_i  ({\bf m})  , \quad \hat{\psi} ({\bf m}) = \sum_i \mathfrak{a}_i \psi_i  ({\bf m}) .
\label{faber}
\end{equation}
Also, the trace of $D^A$ in (\ref{opera}) is given by
\begin{equation}
\mbox{Tr}_\psi \left( D^A  \right) = \sum_i \int_M  \mbox{Tr} \left(  {\psi}_i ,  D^A {\psi}_i\right) .
\label{bbbb}
\end{equation}

 The quantity in (\ref{opera}) is -- up to the trace of $D^A$ -- the principal part of the Hamiltonian for a fermionic quantum field theory of a Weyl spinor \cite{Paschke} for a trivial set of lapse and shift fields $(N,N_\m)$, i.e. $$N=1,\quad N_\m=0,$$ while the quantity in (\ref{NewYork}) involves the Hamilton of a corresponding quantum gauge theory. We shall discuss the connection to bosonic and fermionic quantum field theory in further detail in the next section.

Let us end this section by mentioning that the definition of the Bott-Dirac operator (\ref{BDx22}) involves an additional ambiguity. Let $\lambda\in\mathbb{R}$ and consider the transformation
\begin{eqnarray}
\cb_{\a} \longrightarrow  \cb'_{\a}=    \left(
\begin{array}{cc}
 B_\a  & \lambda C_\a \\ 
\lambda^{-1} \tilde{C}_\a    &  B_\a
\end{array}
\right),
\label{BDx222}
\end{eqnarray}
together with 
\begin{equation}
 \Phi (A) \longrightarrow  \Phi' (A) = \left(
\begin{array}{c}
\cos \left( ({2\tau_2})^{-1}  CS(A)\right)  
\vspace{0,1cm}
\\
\lambda^{-1} \sin\left( ({2\tau_2})^{-1}  CS(A)\right)  
\end{array}
\right)  \exp(-\tau_2^{-1}S(A)).
\nn
\end{equation}
Then $ \Phi' (A)$ lies in the kernel of $\cb_\a'$, i.e.
\begin{equation}
\cb'_{\a} \Phi'(A) = 0.
\nn
\end{equation}
Also, the Hamilton operator (\ref{buddha}) transforms with (\ref{BDx222}) into
\begin{eqnarray}
{\bf H}\longrightarrow {\bf H}=
\left(
\begin{array}{cc}
H_{1}  &  \lambda H^*_{2} \\
 \lambda^{-1} {H}_2 &  {H}_{1} 
\end{array}
\right).
\nn
\end{eqnarray}
Thus, there exist the possibility of varying the relative weight of the two sectors in the direct sum in (\ref{HC}).\\

Note that for ${\bf H}$ to be a good candidate for a Hamilton operator it needs to have a self-adjoint extension. In order to prove this we need to look closer at the Hilbert space. Since there are still some details concerning the construction of the Hilbert space that need to be worked out, such as the convergence question, we leave this question to be addressed in a later publication.

\section{Connection to perturbative quantum field theory}
\label{conqft}

In this section we outline how the Bott-Dirac operator defined in (\ref{BDx22}) together with a representation of the $\mathbf{HD}(M)$ algebra defined in \cite{Aastrup:2012vq,AGnew} and the Hilbert space constructed in \cite{Aastrup:2019yui}, form a candidate for a non-perturbative quantum Yang-Mills theory coupled to a fermionic field on a curved background. This quantum theory involves higher derivative terms, which places it in some proximity to non-local quantum field theory, as well as a topological Yang-Mills term. Finally, we discuss the metric invariant (\ref{bbbb}), which is related to the eta-invariant first introduced by Atiyah, Patodi and Singer \cite{Atiyah}-\cite{AtiyahIII}.
This section is partly based on section 7 in \cite{Aastrup:2019yui}.

\subsection{The bosonic sector}

We will first discuss the bosonic part of ${\bf H}$ in (\ref{buddha}), i.e. the two parts
$$
H_1 \big\vert_{\mbox{\tiny bosonic}} \quad \mbox{and}\quad H_2 \big\vert_{\mbox{\tiny bosonic}}.
$$
Let us first consider the limit $\tau_1=0$ and let us begin by defining formal local field operator $\hat{E}({\bf m})$ as
\begin{equation}
    \hat{E} ({\bf m}) =\sum_j  \xi_j ({\bf m}) \frac{\partial }{\partial \xi_j} .
     \label{vladp}
\end{equation}
In \cite{Aastrup:2019yui} we demonstrated that the formal operators ${A}$ and $\hat{E}$ combine to form the canonical commutation relations for a quantum gauge theory. 
We would like to interpret the expression $\sum_j \frac{\partial}{\partial \xi_j} \frac{\partial}{\partial \xi_j}$ in $H_1$ in (\ref{NewYork}) in terms of the operator $\hat{E}$. To this end we write down the reverse of (\ref{vladp}) 
$$
\frac{\partial}{\partial \xi_j} = \int \xi_j({\bf m})\hat{E}({\bf m}) dm , 
$$
which gives us
\begin{equation} 
\sum_j \frac{\partial}{\partial \xi_j} \frac{\partial}{\partial \xi_j} = \int \hat{E}({\bf m}) \hat{E}({\bf m}) dm , \quad (\tau_1=0)
\label{philipglass}
\end{equation}
where we used $\sum_j \xi_j({\bf m}_1)\xi_j({\bf m}_2)=\delta ({\bf m}_1-{\bf m}_2)$.

Let us now consider what happens when $\tau_1\not=0$. In this case the set $\{ \xi_i\}$ is orthonormal with respect to the Sobolev norm (\ref{inner}) on $\OO^1(M,\mathfrak{g})$, which is constructed via the covariant Laplace operator $\Delta_A$ for a given connection $A\in\ca$. 
Since the eigenvectors $\xi_i$ now depend on $A$ we adopt the notation
\begin{equation}
\hat{E}_A({\bf m})=\sum_j \xi_j({\bf m}) \frac{\partial }{\partial \xi_j} .
\label{skrig}
\end{equation}
Let $\lambda_i$ be the eigenvalues of $\Delta_A$ and $\{e_i\}$ the associated $L^2$-eigenvectors. In the case of the norm (\ref{inner}) we have 
$$
\xi_i=\frac{e_i}{1+\tau_1\lambda_i^p},
$$
where $\{e_i\}$ is an orthonormal basis with respect the the $L^2$-norm, 
but as we did in section \ref{ovn2} we can instead consider any bounded function $f:[0, \infty) \to \R$ with 
$$
\lim_{x\to \infty} f(x)=0,
$$ 
instead of $f(x)=(1+\tau_1x^p)^{-1}$. Then
\begin{equation}
 K_{f,A}({\bf m}_1,{\bf m}_2)= \sum_j \xi_j({\bf m}_1)\xi_j({\bf m}_2)
\label{Kairo}
\end{equation}
is the integral kernel of $f^2(\Delta_A)$, i.e. 
$$f^2(\Delta_A)(\eta)({\bf m}_1)=\int_M K_{f,A} ({\bf m}_1,{\bf m}_2)\eta({\bf m}_2) dm_2 ,$$
which is easily seen by evaluating both sides on the $e_i$'s.

Let us now consider what happens to equation (\ref{philipglass}) when $\tau_1\not=0$. In that case we find the expression
\begin{eqnarray*}
\sum_j \frac{\partial}{\partial \xi_j} \frac{\partial}{\partial \xi_j} 
 & = &\int \int  \hat{E}({\bf m}_1) K_{f,A}({\bf m}_1,{\bf m}_2) \hat{E}({\bf m}_2)dm_1 dm_2 \\
 &=& \int (\hat{E}({\bf m}), f^2 (\Delta_A)(\hat{E})({\bf m}))_{\mbox{\tiny 1-forms}} d m,
\end{eqnarray*}
where $(\cdot , \cdot )_{\mbox{\tiny 1-forms}}$ is again the point-wise inner product on $\OO^1(M,\mathfrak{g})$.
Note that we here have the $\hat{E}$ operators as defined in (\ref{vladp}) and not the $\hat{E}_A$ operators defined in (\ref{skrig}). If we consider the limit $\tau_1\to 0$ there is an expansion in terms of kernels
$$  K_{f,A}({\bf m}_1,{\bf m}_2) \sim k_0({\bf m}_1,{\bf m}_2)+\tau_1^{{1/2p}} k_1({\bf m}_1,{\bf m}_2)+\ldots ,$$
where it is understood that $k_i$ depend on $A$ except for $k_0$, which is 
$$
k_0({\bf m}_1,{\bf m}_2)=\delta({\bf m}_1 - {\bf m}_2)\mathbf{1}_{\mathfrak{g}},
$$ 
where $\mathbf{1}_\mathfrak{g}$
is the identity in the Lie algebra.
Hence we get 
\begin{eqnarray}
\int \int  \hat{E}({\bf m}_1) K_{f,A}({\bf m}_1,{\bf m}_2) \hat{E}({\bf m}_2)dm_1 dm_2 
\hspace{-6cm}&&\nn\\
&\sim& \int \hat{E} ({\bf m}_1) k_0({\bf m}_1,{\bf m}_2)\hat{E} ({\bf m}_2)dm_1dm_2
\nn\\&&
+\tau_1^{{1/2p}} \int \hat{E} ({\bf m}_1) k_1 ({\bf m}_1,{\bf m}_2)\hat{E} ({\bf m}_2) dm_1 dm_2+\ldots   \nn \\
&=&         \int (\hat{E} ({\bf m}) ,\hat{E} ({\bf m}))_{\mbox{\tiny 1-forms}}dm \nn
\\
&& +\tau_1^{{1/2p}} \int (\hat{E} ({\bf m}), 2f'(0)f(0)\Delta_A( \hat{E}) ({\bf m}))_{\mbox{\tiny 1-forms}} dm +\ldots    
\label{Boris}
\end{eqnarray}

The integral kernel also turns up in the two other terms in $H_1\big\vert_{\mbox{\tiny bosonic}}$ in (\ref{NewYork}). We will here only consider the first term that involves the Chern-Simons functional; the other term will give rise to higher-order derivative terms. 
Using (\ref{godhavn}) together with (\ref{Kairo}) we obtain
\begin{eqnarray*}
\sum_i \left(\frac{\pa CS}{\pa \xi_i} \right)^2  =  \int \int  K^{ab}_{f,A}({\bf m}_1,{\bf m}_2) F^a({\bf m}_1) F^b({\bf m}_2)  dm_1 dm_2,
\end{eqnarray*}
which combined with (\ref{Boris}) gives us
\begin{equation}
H_1 \big\vert_{\mbox{\tiny bosonic}} = \int \left(\hat{E}^2   + F^2 \right) + \co(\tau_1^{{1/2p}}) + \mbox{higher derivative terms} ,
\label{Hbb}
\end{equation}
where the higher orders in $\tau_1^{{1/2p}}$ are computed via the integral kernel $K_{f,A}$. We recognize the first term in (\ref{Hbb}) as the Hamilton operator of a Yang-Mills theory.

We would like to emphasise how the covariant UV-regularisation given by the function $f$ in the Sobolev norm (\ref{innerX}) gives rise to higher-derivative interactions in (\ref{Hbb}). From a perturbative point of view the UV-regularisation will simply look like ordinary higher-order interactions and it is only when seen from a non-perturbative point of view that one can fully appreciate the non-local nature of the Sobolev norm.

\begin{remark}
Note that the inclusion of the complex phase to the ground state, that involved the Chern-Simons term in (\ref{first}), is essential for the Hamilton operator (\ref{Hbb}) to emerge. It is interesting that a topological term is required for a Yang-Mills type theory to emerge from this formalism. 
\end{remark}


Let us next write down the commutation relations
$$
\left( \left[ \hat{E}_A({\bf m}_2),{A}({\bf m}_1) \right] \eta\right)(A)  = K_{f,A}({\bf m}_2,{\bf m}_1) \eta(A),
$$
where $A({\bf m})=\sum_i x_i \xi_i({\bf m})$. To lowest order in $\tau_1$ this gives us the canonical commutation relations
\begin{equation}
\left( \left[ \hat{E}_A ({\bf m}_2),{A}({\bf m}_1) \right]\eta\right)(A)  = \d({\bf m}_2-{\bf m}_1) \eta(A) + \co(\tau_1^{{1/2p}}).
\label{Hbbb}
\end{equation} 
Combining (\ref{Hbbb}) with (\ref{Hbb}) shows that in the limit $\tau_1\rightarrow 0$ our formalism coincides with that of a Yang-Mills theory on a curved background. The expansions 
$$
\hat{E}_A ({\bf m})=\sum_i \xi_i({\bf m})\frac{\pa}{\pa \xi_i},\quad A({\bf m})=\sum_i x_i \xi_i({\bf m})
$$ 
correspond in the flat limit to the plane wave expansions of the conjugate fields, which are used in ordinary quantum field theory on a flat manifold. Note here, however, that since the $\xi_i$'s are real the plane wave expansion will be a combination of sines and cosines.

Finally, let us also consider the bosonic term in $H_2\big\vert_{\mbox{\tiny bosonic}}$.  Using (\ref{godhavn})  we find
\begin{equation}
-\frac{1}{2}\sum_j \frac{\pa CS}{\pa \xi_j}   \frac{\pa }{\pa \xi_j}  = \int_M \mbox{Tr} \left( \hat{E}_A \wedge F(A) \right),
\label{topo}
\end{equation}
which looks like a 'BF'-type topological term \cite{Birmingham}. In fact, this term is the Hamiltonian version of a topological Yang-Mills Lagrangian term \cite{Witten:1988ze} 
$$
\cl_{\mbox{\tiny topological YM}} = \int_{\cm}\mbox{Tr} \left( \cf \wedge \cf \right),
$$
where $\cm$ is a four-dimensional manifold and $\cf$ the corresponding four-dimensional field-strength tensor. This suggest that the emergent quantum field theory will have a non-trivial instanton sector. Note that the term (\ref{topo}) is a direct consequence of the ground state involving the Chern-Simons functional.

\subsection{The fermionic sector }

Let us now turn to the fermionic sector given by the terms $H_1\big\vert_{\mbox{\tiny fermionic}}$ and $H_2\big\vert_{\mbox{\tiny fermionic}}$ in ${\bf H}$ in (\ref{Pl1}) and (\ref{Pl2}) as well as the field operators, which we introduced in (\ref{faber}). 

Similar to what we did in the bosonic sector we define the integral kernel
\begin{equation}
\tilde{K}_{f,A}({\bf m}_1,{\bf m}_2)= \sum_i \psi_i({\bf m}_1)\psi^*_i({\bf m}_2),
\label{Capetown}
\end{equation}
where we adopt the general Sobolev norm (\ref{innerspinorX}). Similar to the bosonic sector  we then find
\begin{equation}
\{ \hat{\psi} ({\bf m}_1) ,   \hat{{\psi}}^\dagger ({\bf m}_2) \} = \tilde{K}_{f,A}({\bf m}_1,{\bf m}_2),
\label{Fermionic}
\end{equation}
where again the integral kernel $\tilde{K}_{f,A}({\bf m}_1,{\bf m}_2)$ to lowest order in $\tau_1$ gives the delta function $\d^3({\bf m}_1-{\bf m}_2)$. Thus, we find that the canonical anti-commutation relations emerge from our construction with a correction at order $\tau_1$.

Next consider the Hamiltonian in (\ref{buddha}). Using the Dirac representation of the gamma matrices
$$
\gamma^0 =
 \left(\begin{array}{cc}
\mathds{1}_2 & 0\\
0 & -\mathds{1}_2
\end{array}\right), 
\quad
\gamma^a =
 \left(\begin{array}{cc}
0 &  \sigma^a\\
-  \sigma^a & 0
\end{array}\right),
\quad
\gamma^5=
 \left(\begin{array}{cc}
0 & \mathds{1}_2\\
\mathds{1}_2 & 0
\end{array}\right)
$$
we can rewrite it as
\begin{equation}
{\mathbf{H}} = H_{\mbox{\tiny YM}} \otimes \mathds{1}_2 + H_{\mbox{\tiny Dirac}} + H_{\mbox{\tiny t-YM}} \otimes \left(\begin{array}{cc}
0&1 \\
  1 & 0
\end{array}\right)  + \mbox{higher orders},
\label{vovvov}
\end{equation}
where $H_{\mbox{\tiny YM}}$ is the Yang-Mills Hamiltonian and where $H_{\mbox{\tiny Dirac}}$ has the structure of the Dirac Hamiltonian of a 4-spinor for a pair of trivial lapse and shift fields \cite{Paschke}. Also, $H_{\mbox{\tiny t-YM}}$ is the part related to topological Yang-Mills discussed in the previous section while "higher orders" are terms stemming from either $Q^{ab}$ or $\OO(\bar{c},A)$ in (\ref{b222}).
Note, however, that the structure of the Fock-space in which $H_{\mbox{\tiny Dirac}}$ acts is that of a direct sum and not a product, see (\ref{HC}), which is different from that of ordinary fermionic quantum field theory. Since this appears to be a genuine feature of this formalism the interpretation in (\ref{vovvov}) may be misplaced; what we have now is simply the Hamiltonian of a Weyl spinor in an off-diagonal structure. 

We expect that the Dirac spinors will emerge once we take both representations of $Cl(3)$ into account and not just one of them, as we did in section 3.2 when we constructed the map $\varphi_\chi$.

\subsection{The trace of $D^A$ }

So far we have discussed all the terms in the Hamilton operator ${\bf H}$ in (\ref{buddha}) except one, namely the term $\Tr_\Psi (D^A)$ in equation (\ref{opera}).
This term  is interresting. It is a spectral invariant that measures the spectral asymmetry of the spatial Dirac operator $D^A$. It is related to the eta-invariant for $D^A$ that was first introduced by Atiyah, Patodi and Singer \cite{Atiyah}-\cite{AtiyahIII}. For $\tau_1=0$ $\Tr_\Psi (D^A)$ equals $\eta_{D^A}(-1)$. It is known \cite{AtiyahIII} that $\eta_{D^A}(s)$ has a simple pole at $s=-1$. 

Let us write the trace of $D^A$ in a form that shows the UV-regularisation
\begin{equation}
    \sum_i \langle \psi_i \vert D^A \psi_i \rangle = \sum_i \frac{\lambda_i}{1+\tau_1 (\lambda_i^2)^p},
\end{equation}
where $\{\lambda_i\}$ are the eigenvalues of $D^A$. This expression exist for $p$ big enough. The question is what happens if we remove the regularisation. There are two ways of doing this: if we have
\begin{equation}
     \sum_i \frac{\lambda_i}{1+\tau_1 (\lambda_i^2)^p}
\end{equation}
and take the limit $\tau_1\rightarrow 0$, or if we have
\begin{equation}
     \sum_i \frac{\lambda_i}{(1+\tau_1 (\lambda_i^2))^p}
\end{equation}
and take the limit $p\rightarrow 0$ ($\zeta$-regularisation).
We will only discuss the latter possibility. We note that
\begin{equation}
     \sum_i \frac{\lambda_i}{(1+ (\lambda_i^2))^p}
     = \mbox{Tr} \left( D^A (1+ \Delta_A )^p\right)
\end{equation}
when $p$ is big enough. According to Kontsevich and Vishik, proposition 4.1 in \cite{Kontsevich1} (see also \cite{Kontsevich2}), this function is meromorphic in $p$ and has no pole in $p=0$ and hence 
$$
\lim_{p\rightarrow 0} \sum_i \frac{\lambda_i}{(1+\tau_1 (\lambda_i^2))^p}
$$
makes sense. Also, this expression is independent of the choice $(1+\Delta_A)^p$ in the sense that one can choose other operators than $(1+\Delta_A)$ as long as they fulfill certain properties. 

All this means that the Residue at $s=-1$ of the eta-function $\eta_{D^A}$, which is also known as the Wodzicki trace of $D^A$ \cite{wodzicki}, vanishes, i.e. that $\Tr_\Psi (D^A)$ in (\ref{bbbb}) exist, and, furthermore, that it is independent of the UV-regularisation given by the function $f$ in (\ref{innerX}) and (\ref{innerspinorX}).\\

Note also that:
\begin{itemize}
\item
since $\Tr_\Psi (D^A)$ depends on the gauge field $A$ it will be subject to quantum corrections and hence to time-evolution (see section \ref{sec-time}).
\item
in ordinary fermionic quantum field theory the trace $\Tr_\Psi (D^A)$ corresponds to the supposedly infinite contribution
$$
\int d^3 p E_p \d^{(3)}(0)  
$$
 which is usually removed using normal ordering (see for example \cite{Tong} page 110). 
\end{itemize}
Based in part on the last remark we would like to put forward the speculation that the trace $\Tr_\Psi (D^A)$ may be related to a a time-dependent cosmological constant. More analysis is required, however, to determine whether this could be true.

\section{Time evolution}
\label{sec-time}

The construction of the Hamilton operator ${\bf H}$ in (\ref{buddha}) gives rise to a natural time-evolution $\varphi_t $ of bounded operators in $\mathscr{H}$
\begin{eqnarray}
\varphi_t (\co) = e^{i t {\bf H}} \co e^{- i t {\bf H}} ,\quad \co \in B(\mathscr{H})
\end{eqnarray}
with $\varphi_{t_1}\cdot \varphi_{t_2} = \varphi_{t_1 + t_2}$.

In particular this implies that the metric on $\ca$ constructed in section \ref{ovn2} will have a time evolution since the Sobolev norms in (\ref{inner}) and (\ref{innerX}) are constructed with the covariant Laplace operator which depends on $A$. This, in turn, implies that the regularisation given by the function $f$ in (\ref{innerX}) and the background metric $g$ on $M$ must also have some kind of time-evolution.
Indeed, let us consider the metric (\ref{innerX}) and write it as $\langle \cdot \vert \cdot \rangle_{A,f,g}$ to emphasise also its dependency on the metric. We denote its time-evolution from $t=t_0$ to $t=t_1$ by
$$
\langle \cdot \vert \cdot\rangle_{A,f,g} (t_1) := e^{i (t_1-t_0) {\bf H}} \langle \cdot \vert \cdot\rangle_{A,f,g}  e^{- i (t_1-t_0) {\bf H}}.
$$
and ask the question whether there exist a function $f'$ and a metric $g'$ so that
$$
\langle \cdot \vert \cdot\rangle_{A,f,g} (t_1) \stackrel{?}{=} \langle \cdot \vert \cdot\rangle_{A,f',g'}.
$$
In fact, with the present construction $g'$ will equal $g$ since the metric $\langle \cdot \vert \cdot\rangle_{A,f,g} $ must coincide with the time-evolved metric in the limit $\tau_1=0$. Thus, it is only the function $f$ that evolves.

This could, however, be changed. If, for instance, we choose a metric $g_0$ on $M$ and pick a real number $\tau$, and if we choose the configuration space $\ca$ to be a space of Ashtekar connections \cite{Ashtekar:1986yd,Ashtekar:1987gu}, then we can define
\begin{equation}
g_A := \exp(\tau H_{\mbox{\tiny grav}}) (g_0),
\label{metric-time}
\end{equation}
where $H_{\mbox{\tiny grav}}$ is the gravitational Hamiltonian for a pair of Ashtekar variables $(A,E)$, with $E$ being a densitized triad field compatible with the metric $g$, and where the exponential is with respect to $\{\cdot ,\cdot \}_{\mbox{\tiny P.B.}}$, the Poisson brackets of the Ashtekar variables. With this we can then consider the metric on $\ca$ given by
$\langle \cdot \vert \cdot\rangle_{A,f,g_A}$ that depends on the metric $g_A$. The latter will then be an operator and thus have a time-evolution too.
We can therefore ask the above question again, namely if there exist a pair $(f,g_A)$ so that
$$
\langle \cdot \vert \cdot\rangle_{A,f,g_A} (t_1) = \langle \cdot \vert \cdot\rangle_{A,f',g'_A} (t_0)
$$
holds.
We believe that the answer to this question must be in the affirmative but clearly this remains to be proven. 

Another possibility is to permit a more general class of metrics than what we chose in section 2. This would then entail a more involved time evolution that would also involve the metric $g$.

It is interesting that the non-covariant metric, which we constructed in \cite{Aastrup:2017vrm}, will be static under the time-evolution given by the operator ${\bf H}$. It is the requirement of gauge-covariance that makes the metric on $\ca$ time-dependent.

\section{Discussion}
\label{secdiscussion}

 In this paper we present a geometrical construction {\it over} a configuration space of gauge connections and show that it gives rise to a candidate for a non-perturbative quantum Yang-Mills theory coupled to a fermionic sector on a curved background.
 
Perhaps one of the most surprising features of this construction is that it presents a candidate for a fundamental theory where gravity remains classical. Since the construction includes a gauge-covariant and dynamical ultra-violet regularisation there does not appear to be any need nor any room for a quantisation of gravity. \\

  The construction involves a number of novel concepts, such as gauge covariant metrics on configuration spaces, infinite-dimensional gauge covariant Bott-Dirac operators and gauge covariant, dynamical regularisation, which in turn raises a number of questions. In the following we will discuss some of these questions.

 But before we do that we would like to emphasise that the central object in this construction is the configuration space itself and its geometry. The underlying spatial manifold, its metric and the spatial regularisation are all secondary objects and must be understood in terms of the geometry of the configuration space. The underlying manifold becomes important only when the construction is interpreted in terms of (perturbative) quantum field theory.

One of the technical novelties in this construction is the covariant ultra-violet regularisation. The construction of the Bott-Dirac operators (\ref{F1}) and (\ref{BDx22}) depends on the Sobolev norms (\ref{inner}) and (\ref{innerspinor}) (or more generally (\ref{innerX}) and (\ref{innerspinorX})), which serve as a gauge-covariant ultra-violet regularisation by dampening degrees of freedom below the scale $\tau_1$. This ultra-violet regulator is critical also for the construction of the Hilbert space, see \cite{Aastrup:2019yui}. In an ordinary perturbative quantum field theoretical setup an ultra-violet regularisation of this kind would be interpreted as a non-physical computational artefact, which must ultimately be removed. But the present situation is decidedly different: first of all, the regularisation is part of the metric data on the configuration space $\ca$ and hence it would make little sense -- at least from a mathematical perspective -- to take a singular limit hereof. Secondly, the regularisation depends on $A$ and is therefore itself an operator subjected to time-evolution. This implies that there is no problem of choice since there will be an evolution of regularisations. 

From a perturbative perspective the regularisation will take the form of an infinite series of higher-order interactions. Note that a priori there is no room for a coupling constant between the Yang-Mills and the Dirac Hamiltonians, but in a perturbative framework the regularisation will provide an infinite series of coupling constants, all given by the various derivatives of the function $f$ at zero. Since the regularisation given by the function $f$ has a time-evolution so will the coupling constants. 



On a more general note we find it an interesting possibility that a physically realistic ultra-violet screening can be achieved while gravity remains classical. The usual argument is that distances shorter than the Planck scale must be operational meaningless since a measurement thereof will generate a black hole horizon and hence make observations impossible \cite{Doplicher:1994tu}. It is generally expected that such a screening will be the effect of a quantum theory of gravity but in the present setup an ultra-violet screening is dictated by representation theory.

Another question is related to the choice of trivialisation of $T\ca$. In sections 3 and 4 when we constructed the Bott-Dirac operators we indirectly assumed the existence of a global trivialisation of $T\ca$, which we used to make sense of $\frac{\pa}{\pa\xi_i}$. There is, however, a more canonical choice, which is to use a Levi-Civita connection on $T\ca$.
In this case the Bott-Dirac operator (\ref{F1}) will have the form
$$
B_\chi  =\sum_{ij} b_{ij}(\chi) \left(\tau_2\bar{c}_{i} \nabla_{\xi_j} + {c}_{i} \frac{\pa S}{\pa\xi_j} \right) ,
$$
with
$$
\nabla_{\xi_i}= \frac{\pa}{\pa\xi_i}+ \oo_{\xi_i},
$$
where $\oo$ is a Levi-Civita connection in $T\ca$ that preserves the metric (\ref{inner}) and has vanishing torsion. Note that $B_\chi \Psi (A)=0$ still holds since $\Psi{(A)}$ is a scalar in $T\ca$. 
With the Levi-Civita connection $\oo$ we can write down its curvature 
$
F(\oo)
$, which will show up in the square of the Bott-Dirac operator via the general Bochner identity. It is likely that the curvature $F(\oo)$ will involve the scalar curvature of $M$ and thus give us the Hamiltonian of general relativity.

Yet another interesting question concerns the metric invariant $\Tr_\Psi (D^A)$, which emerges from our framework alongside the Dirac Hamilton operator. We know that this invariant exist and that it is independent of the regularisation, but we do not know what physical interpretation it has. One tentative speculation is that it could be related to a time-dependent cosmological constant.  Note here that $\Tr_\Psi (D^A)$ will involve quantum corrections since it depends on the gauge field $A$. This implies that it will have a time-evolution.

A key conceptual question is that of the choice of configuration space. The key input in this framework is the choice of gauge group. A canonical choice would be that of spatial rotations, i.e. that $\ca$ is a configuration space of spin-connections. This would add a gravitational dimension to the construction and also enable the mechanism we devised in section (\ref{sec-time}) for a time-dependent background metric. Another possibility would be to choose a configuration space that matches the gauge structure of the standard model. Even though this choice is feasible, it is, however, not very appealing since it abandons any hope of explaining the origin of this very gauge structure. 

In fact, one of the original motivations behind this project \cite{Aastrup:2005yk} was to find an explanation for the almost-commutative spectral triple that Chamseddine and Connes has identified in their work on the standard model \cite{Connes:1996gi,Chamseddine:1991qh}. The idea was that an almost-commutative algebra could emerge from an algebra of holonomy-loops in a semi-classical limit. With the present construction we seem to have the key ingredients to complete such an argument and yet there seem to be some conceptual hurdles, which must be overcome before such an interpretation can be plausible. In fact, it seems that the only way such an interpretation can be realistic is if the present framework describes a theory at a higher energy scale, which then gives rise to the algebraic structures found in the standard model in a low-energy limit. Said differently, if such an interpretation is to be taken seriously then it does not seem possible that the Yang-Mills-Dirac theory that we find is directly identifiable with the quantities encountered in the standard model. More work is required, however, to make any definite statements about this.


\vspace{1cm}
\noindent{\bf\large Acknowledgements}\\

\noindent
We would like to thank Peter W. Michor for helpful information on the infinite-dimensional Levi-Civita connection.
JMG would like to express his gratitude to Ilyas Khan, United Kingdom, to entrepreneur Kasper Bloch Gevaldig, Copenhagen, Denmark, and to the engineering company Regnestuen Haukohl \& K\o ppen, Denmark, for their generous financial support. JMG would also like to express his gratitude to the following sponsors:  Frank Jumppanen Andersen, Anders Arnfred, Ria Blanken, Bart De Boeck, Niels Peter Dahl, Niels Giroud, Claus Hansen, Tanina \& Theo Jenk, Simon Kitson, Troels Fjordbak Larsen, Hans-J\o rgen Mogensen, Tero Pulkkinen, Christopher Skak and  Rolf Sleimann for their financial support, as well as all the backers of the 2016 Indiegogo crowdfunding campaign. JMG would also like to express his gratitude to the Institute of Analysis at the Gottfried Wilhelm Leibniz University in Hannover, Germany, for kind hospitality during numerous visits.


\begin{thebibliography}{99}






\bibitem{Connes:1996gi}
A.~Connes,
``Gravity coupled with matter and the foundation of non-commutative
geometry,''
Commun.\ Math.\ Phys.\  {\bf 182} (1996) 155.



\bibitem{Chamseddine:1996zu}
  A.~H.~Chamseddine and A.~Connes,
  ``The Spectral action principle,''
  Commun.\ Math.\ Phys.\  {\bf 186} (1997) 731.



 
\bibitem{Chamseddine:1996rw}
A.~H.~Chamseddine and A.~Connes,
``A universal action formula,''
[arXiv:9606056].




\bibitem{Chamseddine:1991qh}
A.~H.~Chamseddine and A.~Connes,
``Universal formula for noncommutative geometry actions: Unification of
gravity and the standard model,''
Phys.\ Rev.\ Lett.\  {\bf 77} (1996) 4868.








\bibitem{Chamseddine:2006ep}
  A.~H.~Chamseddine, A.~Connes and M.~Marcolli,
  ``Gravity and the standard model with neutrino mixing,''
  [arXiv:0610241].

\bibitem{Chamseddine:2007hz}
  A.~H.~Chamseddine and A.~Connes,
  ``Why the Standard Model,''
  [arXiv:0706.3688].

\bibitem{Chamseddine:2007ia}
  A.~H.~Chamseddine and A.~Connes,
  ``A Dress for SM the Beggar,''
  [arXiv:0706.3690].




\bibitem{Chamseddine:2008zj}
  A.~H.~Chamseddine and A.~Connes,
  ``The Uncanny Precision of the Spectral Action,''
  Commun.\ Math.\ Phys.\  {\bf 293} (2010) 867.



\bibitem{Chamseddine:2012sw}
A.~H.~Chamseddine and A.~Connes,
``Resilience of the Spectral Standard Model,''
JHEP \textbf{09} (2012), 104.



\bibitem{Aastrup:2012vq}
  J.~Aastrup and J.~M.~Grimstrup,
  ``C*-algebras of Holonomy-Diffeomorphisms and Quantum Gravity I,''
  Class.\ Quant.\ Grav.\  {\bf 30} (2013) 085016.



  
 

\bibitem{AGnew}
  J.~Aastrup and J.~M.~Grimstrup,
  ``C*-algebras of Holonomy-Diffeomorphisms and Quantum Gravity II'',
   J.\ Geom.\ Phys.\  {\bf 99} (2016) 10.













\bibitem{Aastrup:2019yui}
J.~Aastrup and J.~M.~Grimstrup,
``Non-perturbative Quantum Field Theory and the Geometry of Functional Spaces,''
[arXiv:1910.01841].




\bibitem{Aastrup:2017atr}
  J.~Aastrup and J.~M.~Grimstrup,
  ``Nonperturbative quantum field theory and noncommutative geometry,''
  J.\ Geom.\ Phys.\  {\bf 145} (2019) 103466.




  
\bibitem{Higson}
N.~Higson and G.~Kasparov, "E-theory and KK-theory for groups which act properly and isometrically on Hilbert space", 
Inventiones Mathematicae, vol. {\bf 144}, issue 1, pp. 23-74.






\bibitem{Aastrup:2017vrm}
  J.~Aastrup and J.~M.~Grimstrup,
  ``Representations of the Quantum Holonomy-Diffeomorphism Algebra,''
  [arXiv:1709.02943].





\bibitem{Atiyah}
M.~F.~Atiyah, V.~K.~Patodi and I.~M.~Singer
"Spectral asymmetry and Riemannian Geometry. I"
Math. Proc. Camb. Philos. Soc., {\bf 77}, (1975), pp. 43-69.


\bibitem{AtiyahII}
M.~F.~Atiyah, V.~K.~Patodi and I.~M.~Singer
"Spectral asymmetry and Riemannian Geometry. II". Math. Proc. Camb. Philos. Soc., {\bf 78}, (1975), pp 405-432 .


\bibitem{AtiyahIII}
M.~F.~Atiyah, V.~K.~Patodi and I.~M.~Singer
"Spectral asymmetry and Riemannian Geometry. III". Math. Proc. Camb. Philos. Soc., {\bf 79}, (1976), pp 71-99.

 


\bibitem{ConnesBook}
A.~Connes,
``Noncommutative Geometry,'' Academic Press, 1994.


\bibitem{1414300}
A.~Connes and M.~Marcolli,
``Noncommutative Geometry, Quantum Fields and Motives,''
 www.alainconnes.org/docs/bookwebfinal.pdf. 


\bibitem{Krajewski:1996se}
T.~Krajewski,
``Classification of finite spectral triples,''
J. Geom. Phys. \textbf{28} (1998), 1-30.


\bibitem{Jureit:2006yt}
J.~H.~Jureit and C.~A.~Stephan,
``On a Classification of Irreducible Almost-Commutative Geometries IV,''
J. Math. Phys. \textbf{49} (2008), 033502.



\bibitem{Jureit:2009ye}
J.~H.~Jureit and C.~A.~Stephan,
``On a Classification of Irreducible Almost-Commutative Geometries V,''
J. Math. Phys. \textbf{50} (2009), 072301.





\bibitem{Bogo}
N.~N.~Bogolyubov, A.~A.~Logunov and A.~I.~Oksak,
"General principles of quantum field theory",
book (1990).



\bibitem{Brunetti}
R.~Brunetti, C.~Dappiaggi, K.~Fredenhagen and J.~Yngvason, 
"Advances in algebraic quantum field theory"
Springer (2015). 


  

\bibitem{Buchholz}
D.~Buchholz and K.~Fredenhagen, "A C*-algebraic approach to interacting quantum field theories",
[arXiv:1902.06062].






\bibitem{Osterwalder:1973dx}
  K.~Osterwalder and R.~Schrader,
  ``Axioms For Euclidean Green's Functions,''
  Commun.\ Math.\ Phys.\  {\bf 31} (1973) 83.


\bibitem{Wightman}
A.~S.~Wightman,  ``HilbertÃ•s sixth problem: Mathematical treatment of the axioms of physics ``, in F.E. Browder (ed.): Mathematical Developments Arising from HilbertÃ•s Problems, Vol. 28:1 of Proc. Symp. Pure Math., Amer. Math. Soc, 1976, pp. 241 - 268.



\bibitem{Haag:1963dh}
  R.~Haag and D.~Kastler,
  ``An Algebraic approach to quantum field theory,''
  J.\ Math.\ Phys.\  {\bf 5} (1964) 848.


\bibitem{Haag:1992hx}
  R.~Haag,
  ``Local quantum physics: Fields, particles, algebras,''
  Berlin, Germany: Springer (1992) 356 p. 





\bibitem{Jacobson:2004rj}
  T.~Jacobson, S.~Liberati and D.~Mattingly,
  ``Astrophysical bounds on Planck suppressed Lorentz violation,''
  Lect.\ Notes Phys.\  {\bf 669} (2005) 101.






\bibitem{Doplicher:1994tu}
  S.~Doplicher, K.~Fredenhagen and J.~E.~Roberts,
  ``The Quantum structure of space-time at the Planck scale and quantum fields,''
  Commun.\ Math.\ Phys.\  {\bf 172} (1995) 187.






\bibitem{Feynman:1981ss}
  R.~P.~Feynman,
  ``The Qualitative Behavior of Yang-Mills Theory in (2+1)-Dimensions,''
  Nucl.\ Phys.\ B {\bf 188} (1981) 479.

 
\bibitem{Singer:1981xw}
  I.~M.~Singer,
  ``The Geometry of the Orbit Space for Nonabelian Gauge Theories. (Talk),''
  Phys.\ Scripta {\bf 24} (1981) 817.


\bibitem{Orland:1996hm}
  P.~Orland,
  ``The Metric on the space of Yang-Mills configurations,''
  [arXiv:9607134].









\bibitem{Witten:1988ze}
E.~Witten,
``Topological Quantum Field Theory,''
Commun. Math. Phys. \textbf{117} (1988), 353.





\bibitem{michor}
P.~Michor, "Topics in Differential Geometry", Am. Math. Soc. (2008).







\bibitem{Kodama:1988yf}
  H.~Kodama,
  ``Specialization of Ashtekar's Formalism to Bianchi Cosmology,''
  Prog.\ Theor.\ Phys.\  {\bf 80} (1988) 1024.



\bibitem{Smolin:2002sz}
  L.~Smolin,
  ``Quantum gravity with a positive cosmological constant,''
  [arXiv:0209079].




\bibitem{Ashtekar:1986yd}
  A.~Ashtekar,
  ``New Variables for Classical and Quantum Gravity,''
  Phys.\ Rev.\ Lett.\  {\bf 57} (1986) 2244.

\bibitem{Ashtekar:1987gu}
  A.~Ashtekar,
  ``New Hamiltonian Formulation of general relativity,''
  Phys.\ Rev.\  D {\bf 36} (1987) 1587.













\bibitem{Paschke}
M. Paschke and T. Kopf.
"A spectral quadruple for de Sitter space",
J. Math. Phys. {\bf 43}, 818 (2002).






\bibitem{Birmingham}
D.~Birmingham, M.~Blau and G.~Thompson "Geometry and quantization of topological gauge theories". Internat. J. Modern Phys. {\bf A 5} (1990), no. 24, 47214752.








\bibitem{Kontsevich1}
M.~Kontsevich and S.~Vishik,
``Determinants of elliptic pseudo-differential operators,''
Preprint, Max Planck Institut f\"{u}r Mathematics, Bonn (1994).



\bibitem{Kontsevich2}
M.~Kontsevich and S.~Vishik,
``Geometry of determinants of elliptic operators,''
 In: Gindikin S., Lepowsky J., Wilson R.L. (eds) Functional Analysis on the Eve of the 21st Century. Progress in Mathematics (1995), vol 131/132. Birkh\"{a}user Boston.




\bibitem{wodzicki}
M.~Wodzicki. "Spectral Asymmetry and Noncommutative Residue". Thesis, Stekhlov Institute
of Mathematics, Moscow, 1984.







  
\bibitem{Tong}
   D.~Tong,
   "Lecture Notes on Quantum Field Theory."
http://www.damtp.cam.ac.uk/user/tong/qft.html.









\bibitem{Aastrup:2014ppa}
  J.~Aastrup and J.~M.~Grimstrup,
  ``The quantum holonomy-diffeomorphism algebra and quantum gravity,''
  Int.\ J.\ Mod.\ Phys.\ A {\bf 31} (2016) no.10,  1650048.




\bibitem{Aastrup:2015gba}
J.~Aastrup and J.~M.~Grimstrup,
``Quantum Holonomy Theory,''
Fortsch. Phys. \textbf{64} (2016) no.10, 783-818.


























  






\bibitem{Unruh:1976db}
W.~G.~Unruh,
``Notes on black hole evaporation,''
Phys. Rev. D \textbf{14} (1976), 870.





\bibitem{Aastrup:2005yk}
J.~Aastrup and J.~M.~Grimstrup,
``Spectral triples of holonomy loops,''
Commun. Math. Phys. \textbf{264} (2006), 657-681.







\end{thebibliography}
\end{document}